\documentclass{article}

\usepackage{microtype}
\usepackage{graphicx}
\usepackage{subcaption}
\usepackage{booktabs}
\usepackage{CJKutf8}

\usepackage{hyperref}

\usepackage[accepted]{icml2026}

\usepackage{amsmath}
\usepackage{amssymb}
\usepackage{mathtools}
\usepackage{amsthm}

\newenvironment{compactenum}{
  \begin{list}{\arabic{enumi}.}{
    \usecounter{enumi}
    \setlength{\topsep}{4pt plus 1pt minus 2pt}
    \setlength{\partopsep}{1pt plus 0.5pt minus 0.5pt}
    \setlength{\itemsep}{1.5pt plus 1pt minus 0.5pt}
    \setlength{\parsep}{2pt plus 1pt minus 0.5pt}
    \setlength{\leftmargin}{2em}
    \setlength{\labelsep}{5pt}
    \setlength{\labelwidth}{\dimexpr 2em - 5pt\relax}}
}{\end{list}}

\usepackage{worldflags}
\newcommand{\cnflag}{\,\worldflag[width=1.2ex,framewidth=0.05pt]{CN}}
\newcommand{\usflag}{\,\worldflag[width=1.2ex,framewidth=0.05pt]{US}}

\usepackage[capitalize,noabbrev]{cleveref}

\theoremstyle{plain}

\theoremstyle{definition}

\theoremstyle{remark}

\usepackage[disable,textsize=tiny]{todonotes}

\icmltitlerunning{Government AI Use as a Monitoring Primitive}

\begin{document}

\twocolumn[
  \icmltitle{Government AI Use as a Monitoring Primitive: A Public Document Pilot Study}

  \icmlsetsymbol{equal}{*}

  \begin{icmlauthorlist}
    \icmlauthor{David I. Atkinson}{neu}
    \icmlauthor{Joan Eleanor O'Bryan}{harvard}
  \end{icmlauthorlist}

  \icmlaffiliation{neu}{Northeastern University}
  \icmlaffiliation{harvard}{Harvard University}

  \icmlcorrespondingauthor{David Atkinson}{atkinson.da@northeastern.edu}

  \icmlkeywords{AI governance, language models, government AI adoption, AI-text detection, monitoring}

  \vskip 0.3in
]

\printAffiliationsAndNotice{}
\begin{abstract}
\noindent 
Governments are important actors in frontier AI governance, but many facts about their adoption and use of AI systems are difficult to observe directly. Procurement disclosures and official statements are useful, but can also be delayed, selective, and better suited to measuring formal adoption than actual day-to-day use. We propose a complementary monitoring primitive: measuring traces of language-model assistance in public government documents. The approach is lightweight, externally reproducible, and based on revealed behavior rather than stated intent. In a pilot study of ten public document streams from U.S.\ and PRC government-related sources, we find that, while 2021 baselines are consistently near zero, by 2026, four of our ten sources show statistically significant signs of AI-assisted writing. In our sample, the U.S.\ signal concentrates in publications downstream of policy work; the PRC signal concentrates closer to it. We close by discussing how this signal could complement existing instruments for monitoring government AI adoption, and where it falls short.
\end{abstract}

\begin{figure}[t]
    \centering
    \includegraphics[width=\columnwidth]{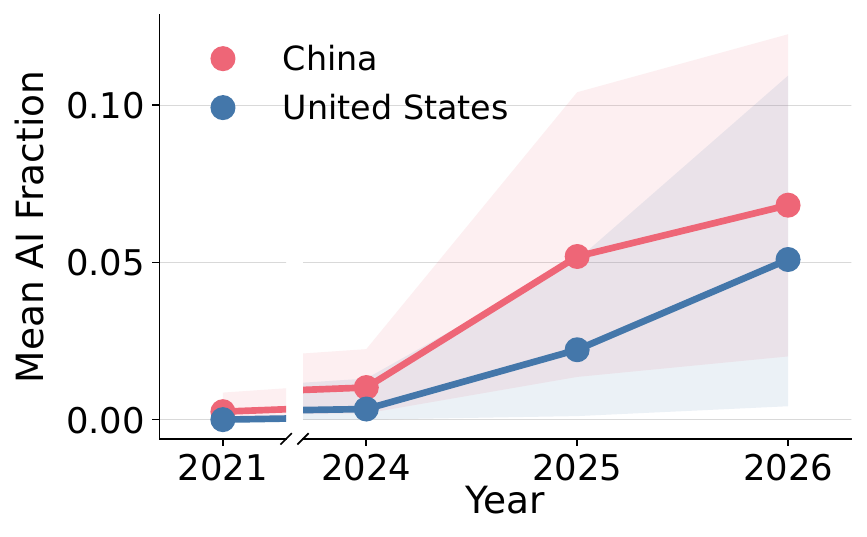}
    \caption{\textbf{Pooled AI-writing signal by country, 2021--2026}. Each marker is the mean per-document fraction of text flagged as AI-generated by Pangram (range 0--1), pooled across each country's sources (Figure~\ref{fig:fraction_per_country_ai_by_year} shows the per-source breakdown). Sources are weighted equally, and documents equally within a source; shaded regions show 95\% two-stage cluster-bootstrap CIs. Both countries sit near zero in 2021 and rise from 2024 on.}
    \label{fig:fraction_country_ai_by_year}
\end{figure}

In 1954, the economist Armen Alchian famously used publicly traded equity prices to infer that lithium was the fusion fuel in the newly developed hydrogen bomb~\citep{newhardStockMarketSpeaks2014}.
In doing so, he neatly illustrated a more general principle: strategically important facts that are not directly observable often leave indirect traces in public data.

\begin{figure*}[!t]
    \centering
    \includegraphics[width=\textwidth]{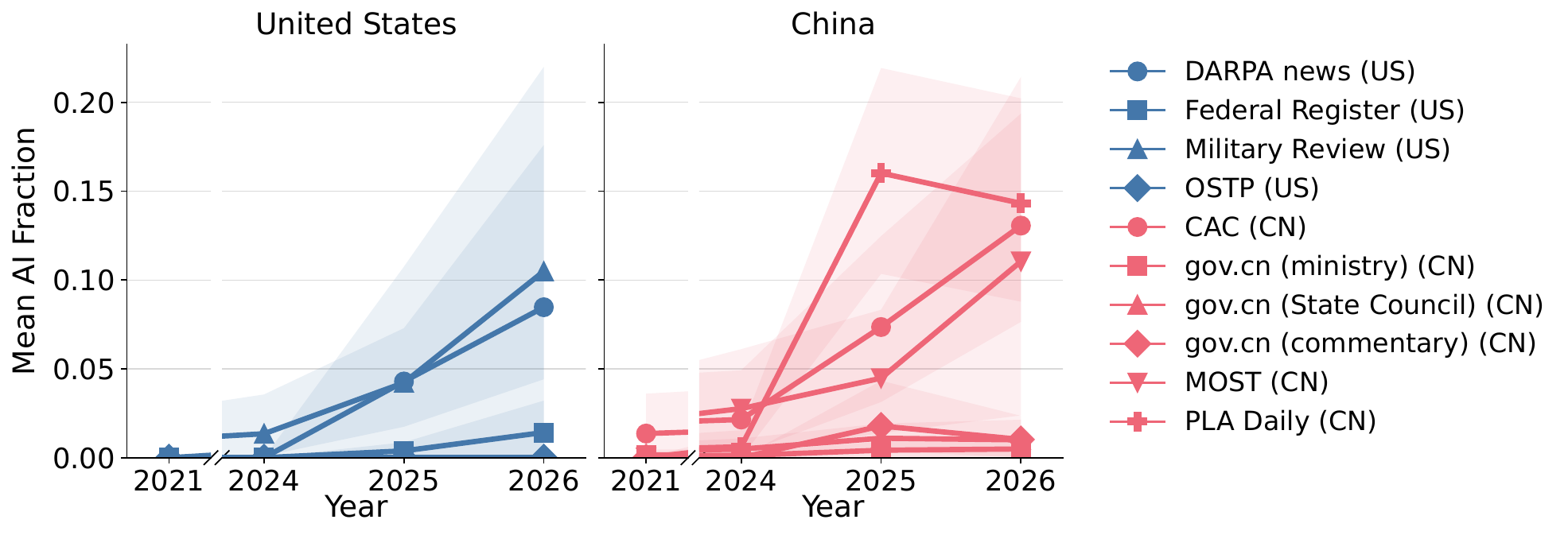}
    \caption{Mean fraction of each document flagged as AI-generated by Pangram, for each source across time. Shaded bands show 95\% bootstrap confidence intervals. All sources show near-zero baselines in 2021; both countries exhibit rising scores from 2024 onward, with Chinese sources reaching a source-weighted pooled mean of 0.07 [95\% CI: 0.02, 0.12] by 2026 and U.S.\ sources 0.05 [0.01, 0.11].}
    \label{fig:fraction_per_country_ai_by_year}
\end{figure*}

One of these strategically important facts is the degree to which governments are familiar with, and using, frontier AI models.
Will governments use frontier AI systems for analysis, propaganda, military planning, cyber operations, or regulatory enforcement?
Will they restrict their own use, subsidize deployment, or quietly integrate models into routine bureaucratic work?
Better answers to these questions would bear on a wide range of proposed governance interventions.
AI use may also be informative about attitudes toward AI itself: experimental work on civilians shows that exposure to LLMs causally shifts policy preferences~\citep{haslbergerRageMachineGenerative2025}, and survey work finds (intuitively) that using AI predicts attitudes toward its use~\citep{chenWhenResearchersUse2026}.\footnote{A fuller discussion of related work is deferred to \cref{app:related_work}.}

The instruments we currently rely on for these questions are valuable but partial.
Public strategies and policy statements describe official priorities, but
can lag practice and emphasize politically salient uses~\citep{gao2025generativeAI,blomquistRacingRecognitionPrestige2026}.
Procurement records, staffing patterns, budget lines, and agency announcements add detail, but are
noisy and difficult to compare across countries~\citep{Johnson_2025,haagStateAICompetition2025}.

We therefore propose an additional signal: traces of language-model assistance in the documents that governments publish in the course of their work. This signal is cheap to collect, externally reproducible, and---importantly---reflects revealed behavior rather than stated intent. We call it a monitoring \emph{primitive} because it is a small, standardized measurement---computed the same way on any public document stream---that can be tracked over time and composed with other instruments into a broader monitoring regime.

\paragraph{Our Contributions}
\begin{compactenum}
    \item We frame public-document AI detection as an ecosystem-monitoring primitive for AI governance.
    \item We report a pilot study comparing document streams from selected U.S.\ government, PRC government, and CCP-related outlets in 2024--2026, with 2021 as a pre-LLM baseline.
    \item We discuss how our proposed signal could fit into a broader monitoring regime, including its limitations, risks, and open technical problems.
\end{compactenum}

\section{Data and Method}
We collect over 3,000 public documents across ten different sources.\footnote{See \cref{app:source_descriptions} for full source descriptions and scraping procedures; \cref{tab:source_coverage} for document counts; \cref{app:random_examples} for sample documents; and \cref{app:data-quality-check} for a spot-check of the ingestion pipeline.} We set 2021 as a pre-mass-market LLM baseline and compare against documents from 2024--2026.
For detection, we use Pangram~\citep{emiTechnicalReportPangram2024}, a commercial AI-text classifier.  Pangram segments each document into windows, classifies each window, and returns a document-level \texttt{fraction\_ai} \(\in [0, 1]\).
We use Pangram because independent evaluations identify it as the strongest available commercial detector, remaining robust to paraphrasing and humanization edits, and matching expert human annotator performance~\citep{jabarianArtificialWritingAutomated2025,russellPeopleWhoFrequently2025}.

\Cref{fig:fraction_country_ai_by_year} presents our top-line results. Baseline rates in 2021 are near zero across all sources. Both countries' pooled scores rise from 2024 onward, reaching a 2026 Chinese pooled mean of 0.07 [0.02, 0.12] and a U.S.\ pooled mean of 0.05 [0.01, 0.11]. 

Source-level patterns are also suggestive.
Among U.S.\ sources, the streams with the highest 2026 scores---\usflag\textit{Military Review} (0.10 [0.04, 0.17]) and \usflag DARPA news (0.08 [0.00, 0.20])---sit downstream of high-level policy generation:
the former
publishes essays on Army doctrine and operations~\citep{armyupressMilitaryReview};
the latter consists largely of public-affairs writeups of completed research programs~\citep{darpaPublicAffairs}.
The U.S.\ sources closest to AI policy formulation in our sample---\usflag OSTP releases and AI-keyworded \usflag Federal Register documents---show no signal at all.

The PRC pattern is closer to the reverse.
Two of the strongest PRC scores come from \cnflag CAC (0.13 [0.07, 0.19]),
the regulator that writes many of the rules under which generative-AI services in China operate~\citep{sheehanChinaAIRegulations2023},
and \cnflag MOST (0.11 [0.02, 0.21]),
the ministry that administers national science-and-technology policy and funding and that, since a 2023 restructuring, serves as the administrative office of the Party's Central Science and Technology Commission~\citep{naughtonReorganizationChinasScience2023}.

\begin{table*}[t]
    \centering
    \resizebox{\textwidth}{!}{\begin{tabular}{llrrrrr}
\toprule
 &  & \multicolumn{4}{c}{Count} &  \\
\cmidrule(lr){3-6}
Source & Description & 2021 & 2024 & 2025 & 2026 & Total \\
\midrule
CAC (CN) & PRC central internet/data/algorithm regulator & 100 & 100 & 100 & 100 & 400 \\
gov.cn (ministry) (CN) & AI-keyworded ministry-issued docs on gov.cn & 100 & 100 & 100 & 65 & 365 \\
gov.cn (State Council) (CN) & AI-keyworded State Council normative instruments & 68 & 50 & 48 & 19 & 185 \\
gov.cn (commentary) (CN) & AI-keyworded official policy interpretations & 100 & 100 & 100 & 100 & 400 \\
MOST (CN) & PRC Ministry of Science and Technology work briefings & 98 & 98 & 100 & 42 & 338 \\
PLA Daily (CN) & Official PLA newspaper, commentary articles & 29 & 100 & 100 & 100 & 329 \\
DARPA news (US) & U.S. DARPA programme/research news & 94 & 64 & 46 & 13 & 217 \\
Federal Register (US) & AI-keyworded U.S.\ Federal Register documents & 100 & 100 & 100 & 86 & 386 \\
Military Review (US) & U.S.\ Army University Press journal essays & 100 & 100 & 100 & 35 & 335 \\
OSTP (US) & White House Office of Science \& Technology Policy news & 75 & 20 & 13 & 5 & 113 \\
\midrule
Total &  & 864 & 832 & 807 & 565 & 3068 \\
\bottomrule
\end{tabular}
}
    \caption{Number of documents collected per source and year. The 2026 column is partial, with a collection cutoff of 2026-04-25.}
    \label{tab:source_coverage}
\end{table*}

\section{Discussion}

Public document detection traces can help answer questions that matter for AI governance but are difficult to study directly. Which parts of the state adopt AI first? Does AI use cluster near militarily or politically sensitive functions, or around more routine administrative activities? Our pilot study begins to shed some light on these and similar questions.

For instance, how do export controls impact the diffusion of frontier models? Because
export-control and licensing regimes increasingly meter access to frontier models and compute~\citep{sutterExportControlsChinaSemiconductors2025,kahnAnthropicDisablesFableMythos2026},
the signal can speak to whether such controls slow government adoption or only commercial adoption, especially in a world where
states may retain privileged access to capabilities the public cannot obtain~\citep{anthropicClaudeGovModels2025}.

Public-document detection can also aid outside verification of the accuracy of government disclosure.
U.S.\ agencies are now required to publish AI use case
inventories~\citep{omb2025inventory}. However, this requirement exempts the Department of
Defense, the Intelligence Community, and national-security
uses~\citep{nextgov2026inventory}---often the areas where external interest is greatest.
Here, revealed-behavior can surface activity that
self-reported inventory omits, and, as per-model attribution abilities mature ~\citep{antoun-etal-2024-text},
could further indicate whether a state's published text reflects domestic versus 
foreign, or closed versus open, model use.

These questions can be difficult to answer in other ways because existing public-sector AI monitoring instruments are limited.
In addition to the use-case restrictions discussed above, audits have found agency inventories to be incomplete or inaccurate~\citep{gaoAIImplementation2023}.
More broadly, instruments such as procurement records, vendor agreements, usage logs, and direct disclosure often require insider access or government self-report. Public-text detection, by contrast, can be run by journalists, academics, NGOs, or other states using only public outputs. This shifts at least one monitoring capability from governments to civil society, and makes it less dependent on the monitored party's own disclosure practices.

Our approach has limitations. AI-text detectors are brittle: they can be vulnerable to deliberate evasion, paraphrasing, and domain shift~\citep{sadasivan2025aigeneratedtextreliablydetected,mady2026featureaugmentedtransformersrobustaitext}.
Detected AI assistance might say more about bureaucratic document production than about the high-level policymakers whose use one most wants to measure. Cross-country comparisons can be confounded by differences in document genre, publication norms, and language-specific differences in detection ability~\citep{liangGPTDetectorsAre2023}. And although Pangram self-reports strong Chinese language performance~\citep{emiTechnicalReportPangram2024}, independent validation of this is limited.

Despite these limitations, the signal is useful precisely because it is cheap, repeatable, and composable. A monitoring organization could run the analysis each month on a fixed panel of sources, with pre-registered selection rules and stable baselines. Changes in detector scores would not by themselves justify strong conclusions, but they could prioritize sources for closer qualitative review, procurement searches, interviews, or comparison with disclosed AI policies.

Several technical improvements would make the approach more useful. The most important is the ability to reliably identify particular models, or model families. This matters in particular if government document streams tend to reflect domestic model use. Other useful improvements include multilingual calibration, perturbation robustness, and methods to distinguish drafting from editing or summarization.

Finally, there is also an important governance question concerning potential negative effects of this kind of monitoring. Public analysis can improve accountability, support open-source intelligence, and help researchers to compare government behavior across jurisdictions. But it could also encourage countermeasures, generate misleading accusations, and contribute to race dynamics if read as evidence that rivals are accelerating adoption. Public reporting should therefore emphasize uncertainty, aggregate trends, and corroboration over confident document-level claims.

\section*{Acknowledgements}
We would like to thank David Bau and the anonymous reviewers at the ICML 2026 Workshop on Technical AI Governance Research (TAIGR) for their suggestions. This work was supported by grants from Pangram and BlueDot Impact's Rapid Grants program. David Atkinson is supported by the National Science Foundation CISE Graduate Fellowship under Grant No. 2313998. Any opinions, findings, and conclusions or recommendations expressed in this material are those of the authors and do not necessarily reflect the views of the National Science Foundation.

\bibliography{references}

\begin{thebibliography}{44}
\providecommand{\natexlab}[1]{#1}
\providecommand{\url}[1]{\texttt{#1}}
\expandafter\ifx\csname urlstyle\endcsname\relax
  \providecommand{\doi}[1]{doi: #1}\else
  \providecommand{\doi}{doi: \begingroup \urlstyle{rm}\Url}\fi

\bibitem[Ansari et~al.(2026)Ansari, Zhang, Tripto, and
  Lee]{ansariEchoesAutomationIncreasing2026}
Ansari, A., Zhang, D.~C., Tripto, N.~I., and Lee, D.
\newblock Echoes of {{Automation}}: {{The Increasing Use}} of {{LLMs}} in
  {{Newsmaking}}, April 2026.

\bibitem[{Anthropic}(2025)]{anthropicClaudeGovModels2025}
{Anthropic}.
\newblock Claude {Gov} models for {U.S.} national security customers.
\newblock Anthropic News, June 2025.
\newblock URL
  \url{https://www.anthropic.com/news/claude-gov-models-for-u-s-national-security-customers}.

\bibitem[Antoun et~al.(2024)Antoun, Sagot, and Seddah]{antoun-etal-2024-text}
Antoun, W., Sagot, B., and Seddah, D.
\newblock From text to source: Results in detecting large language
  model-generated content.
\newblock In Calzolari, N., Kan, M.-Y., Hoste, V., Lenci, A., Sakti, S., and
  Xue, N. (eds.), \emph{Proceedings of the 2024 Joint International Conference
  on Computational Linguistics, Language Resources and Evaluation (LREC-COLING
  2024)}, pp.\  7531--7543, Torino, Italia, May 2024. ELRA and ICCL.
\newblock URL \url{https://aclanthology.org/2024.lrec-main.665/}.

\bibitem[Appel et~al.(2025)Appel, McCrory, Tamkin, McCain, Neylon, and
  Stern]{appelAnthropicEconomicIndex2025}
Appel, R., McCrory, P., Tamkin, A., McCain, M., Neylon, T., and Stern, M.
\newblock Anthropic {{Economic Index}} report: {{Uneven}} geographic and
  enterprise {{AI}} adoption, November 2025.

\bibitem[{Army University Press}(n.d.)]{armyupressMilitaryReview}
{Army University Press}.
\newblock Military review: The professional journal of the {U.S.} army.
\newblock Army University Press, U.S. Army Combined Arms Center, Fort
  Leavenworth, KS, n.d.
\newblock URL \url{https://www.armyupress.army.mil/Military-Review/}.

\bibitem[Blomquist(2026)]{blomquistRacingRecognitionPrestige2026}
Blomquist, K.
\newblock Racing for recognition? {Theorizing} emerging status hierarchies and
  prestige competition in the {AI} era.
\newblock \emph{International Affairs}, 102\penalty0 (3):\penalty0 949--970,
  2026.
\newblock \doi{10.1093/ia/iiag028}.
\newblock URL \url{https://doi.org/10.1093/ia/iiag028}.

\bibitem[Chen \& Jia(2026)Chen and Jia]{chenWhenResearchersUse2026}
Chen, C. and Jia, X.
\newblock When researchers use {{AI}}: Public trust, ethical judgments, and the
  perceived value of academic research.
\newblock \emph{AI and Ethics}, 6\penalty0 (2):\penalty0 223, March 2026.
\newblock ISSN 2730-5961.
\newblock \doi{10.1007/s43681-026-01039-w}.

\bibitem[Cheung \& Lau(2025)Cheung and Lau]{cheungDeepSeekUsePRC2025}
Cheung, S. and Lau, K.-s.
\newblock {{DeepSeek Use}} in {{PRC Military}} and {{Public Security Systems}}.
\newblock \emph{China Brief} 25(20), The Jamestown Foundation, October 2025.
\newblock URL
  \url{https://jamestown.org/deepseek-use-in-prc-military-and-public-security-systems/}.

\bibitem[{Defense Advanced Research Projects Agency}(n.d.)]{darpaPublicAffairs}
{Defense Advanced Research Projects Agency}.
\newblock Communications and public affairs.
\newblock \url{https://www.darpa.mil/news/public-affairs}, n.d.

\bibitem[Denain(2026)]{jsdAddingClinicalTrial2026}
Denain, J.-S.
\newblock Adding {{Clinical}} trial registry entries, {{Earnings}} calls
  prepared remarks and {{Federal}} court opinions from {{CourtListener}}.
\newblock X (formerly Twitter), Epoch AI, March 2026.
\newblock URL \url{https://x.com/i/status/2037646276251279674}.

\bibitem[Emi \& Spero(2024)Emi and Spero]{emiTechnicalReportPangram2024}
Emi, B. and Spero, M.
\newblock Technical {{Report}} on the {{Pangram AI-Generated Text Classifier}},
  July 2024.

\bibitem[{Executive Office of the
  President}(2020)]{PromotingUseTrustworthy2020}
{Executive Office of the President}.
\newblock Promoting the {{Use}} of {{Trustworthy Artificial Intelligence}} in
  the {{Federal Government}}.
\newblock Executive Order 13960, Federal Register, document 2020-27065,
  December 2020.
\newblock URL
  \url{https://www.federalregister.gov/documents/2020/12/08/2020-27065/promoting-the-use-of-trustworthy-artificial-intelligence-in-the-federal-government}.
\newblock Published December 8, 2020.

\bibitem[Haag(2025)]{haagStateAICompetition2025}
Haag, A.
\newblock The state of {AI} competition in advanced economies.
\newblock Feds notes, Board of Governors of the Federal Reserve System,
  Washington, D.C., October 2025.
\newblock URL
  \url{https://www.federalreserve.gov/econres/notes/feds-notes/the-state-of-ai-competition-in-advanced-economies-20251006.html}.

\bibitem[Hans et~al.(2024)Hans, Schwarzschild, Cherepanova, Kazemi, Saha,
  Goldblum, Geiping, and Goldstein]{hansSpottingLLMsBinoculars2024}
Hans, A., Schwarzschild, A., Cherepanova, V., Kazemi, H., Saha, A., Goldblum,
  M., Geiping, J., and Goldstein, T.
\newblock Spotting {{LLMs With Binoculars}}: {{Zero-Shot Detection}} of
  {{Machine-Generated Text}}, October 2024.

\bibitem[Haslberger et~al.(2025)Haslberger, Gingrich, and
  Bhatia]{haslbergerRageMachineGenerative2025}
Haslberger, M., Gingrich, J., and Bhatia, J.
\newblock Rage against the machine? {{Generative AI}} exposure, subjective
  risk, and policy preferences.
\newblock \emph{Journal of European Public Policy}, September 2025.
\newblock \doi{10.1080/13501763.2025.2554903}.
\newblock Advance online publication, September 10, 2025.

\bibitem[Ho et~al.(2024)Ho, Besiroglu, Erdil, Owen, Rahman, Guo, Atkinson,
  Thompson, and Sevilla]{hoAlgorithmicProgressLanguage2024}
Ho, A., Besiroglu, T., Erdil, E., Owen, D., Rahman, R., Guo, Z.~C., Atkinson,
  D., Thompson, N., and Sevilla, J.
\newblock Algorithmic progress in language models, March 2024.

\bibitem[Jabarian \& Imas(2025)Jabarian and
  Imas]{jabarianArtificialWritingAutomated2025}
Jabarian, B. and Imas, A.
\newblock Artificial writing and automated detection.
\newblock Working Paper 34223, National Bureau of Economic Research, 2025.
\newblock URL \url{https://www.nber.org/papers/w34223}.

\bibitem[Johnson et~al.(2025)Johnson, Silva, Leon, Eslami, Schwanke, Dotan, and
  Heidari]{Johnson_2025}
Johnson, N., Silva, E., Leon, H., Eslami, M., Schwanke, B., Dotan, R., and
  Heidari, H.
\newblock Legacy procurement practices shape how u.s. cities govern ai:
  Understanding government employees’ practices, challenges, and needs.
\newblock In \emph{Proceedings of the 2025 ACM Conference on Fairness,
  Accountability, and Transparency}, FAccT ’25, pp.\  772–789. ACM, 2025.
\newblock \doi{10.1145/3715275.3732049}.
\newblock URL \url{http://dx.doi.org/10.1145/3715275.3732049}.

\bibitem[Kahn(2026)]{kahnAnthropicDisablesFableMythos2026}
Kahn, J.
\newblock Anthropic disables fable and mythos ai models after u.s. government
  bars it from giving foreigners access.
\newblock Fortune, June 2026.
\newblock URL
  \url{https://fortune.com/2026/06/13/anthropic-disables-fable-mythos-export-controls-national-security-threat/}.

\bibitem[Kelley(2026)]{nextgov2026inventory}
Kelley, A.
\newblock Agencies report over 3,000 {AI} use cases in 2025.
\newblock Nextgov/FCW, April 2026.
\newblock URL
  \url{https://www.nextgov.com/artificial-intelligence/2026/04/agencies-report-over-3000-ai-use-cases-2025/412898/}.
\newblock Published April 16, 2026. Reports that the Intelligence Community and
  Department of Defense are exempt from inventory reporting.

\bibitem[Kirchenbauer et~al.(2024)Kirchenbauer, Geiping, Wen, Katz, Miers, and
  Goldstein]{kirchenbauerWatermarkLargeLanguage2024}
Kirchenbauer, J., Geiping, J., Wen, Y., Katz, J., Miers, I., and Goldstein, T.
\newblock A {{Watermark}} for {{Large Language Models}}, May 2024.

\bibitem[Liang et~al.(2023)Liang, Yuksekgonul, Mao, Wu, and
  Zou]{liangGPTDetectorsAre2023}
Liang, W., Yuksekgonul, M., Mao, Y., Wu, E., and Zou, J.
\newblock {{GPT}} detectors are biased against non-native {{English}} writers,
  July 2023.

\bibitem[Liang et~al.(2024)Liang, Izzo, Zhang, Lepp, Cao, Zhao, Chen, Ye, Liu,
  Huang, McFarland, and Zou]{liangMonitoringAIModifiedContent2024}
Liang, W., Izzo, Z., Zhang, Y., Lepp, H., Cao, H., Zhao, X., Chen, L., Ye, H.,
  Liu, S., Huang, Z., McFarland, D.~A., and Zou, J.~Y.
\newblock Monitoring {{AI-Modified Content}} at {{Scale}}: {{A Case Study}} on
  the {{Impact}} of {{ChatGPT}} on {{AI Conference Peer Reviews}}, March 2024.

\bibitem[Liang et~al.(2025)Liang, Zhang, Wu, Lepp, Ji, Zhao, Cao, Liu, He,
  Huang, Yang, Potts, Manning, and Zou]{liangQuantifyingLargeLanguage2025}
Liang, W., Zhang, Y., Wu, Z., Lepp, H., Ji, W., Zhao, X., Cao, H., Liu, S., He,
  S., Huang, Z., Yang, D., Potts, C., Manning, C.~D., and Zou, J.
\newblock Quantifying large language model usage in scientific papers.
\newblock \emph{Nature Human Behaviour}, 9\penalty0 (12):\penalty0 2599--2609,
  December 2025.
\newblock ISSN 2397-3374.
\newblock \doi{10.1038/s41562-025-02273-8}.

\bibitem[Mady et~al.(2026)Mady, Reschke, and
  Schuller]{mady2026featureaugmentedtransformersrobustaitext}
Mady, M., Reschke, J., and Schuller, B.
\newblock Feature-augmented transformers for robust ai-text detection across
  domains and generators, 2026.
\newblock URL \url{https://arxiv.org/abs/2605.03969}.

\bibitem[Misra et~al.(2025)Misra, Wang, McCullers, White, and
  Ferres]{misraMeasuringAIDiffusion2025}
Misra, A., Wang, J., McCullers, S., White, K., and Ferres, J.~L.
\newblock Measuring {{AI Diffusion}}: {{A Population-Normalized Metric}} for
  {{Tracking Global AI Usage}}, November 2025.

\bibitem[Mitchell et~al.(2023)Mitchell, Lee, Khazatsky, Manning, and
  Finn]{mitchellDetectGPTZeroShotMachineGenerated2023}
Mitchell, E., Lee, Y., Khazatsky, A., Manning, C.~D., and Finn, C.
\newblock {{DetectGPT}}: {{Zero-Shot Machine-Generated Text Detection}} using
  {{Probability Curvature}}, July 2023.

\bibitem[Naughton et~al.(2023)Naughton, Cheung, Xiao, Xu, and
  Yang]{naughtonReorganizationChinasScience2023}
Naughton, B., Cheung, T.~M., Xiao, S., Xu, Y., and Yang, Y.
\newblock Reorganization of {China}'s science and technology system.
\newblock IGCC Working Paper, UC Institute on Global Conflict and Cooperation,
  2023.
\newblock URL
  \url{https://ucigcc.org/publication/reorganization-of-chinas-science-and-technology-system/}.

\bibitem[Newhard(2014)]{newhardStockMarketSpeaks2014}
Newhard, J.~M.
\newblock The stock market speaks: {{How Dr}}. {{Alchian}} learned to build the
  bomb.
\newblock \emph{Journal of Corporate Finance}, 27:\penalty0 116--132, August
  2014.
\newblock ISSN 0929-1199.
\newblock \doi{10.1016/j.jcorpfin.2014.05.002}.

\bibitem[{Office of Management and Budget}(2025)]{omb2025inventory}
{Office of Management and Budget}.
\newblock 2025 federal agency artificial intelligence use case inventory.
\newblock
  \url{https://github.com/ombegov/2025-Federal-Agency-AI-Use-Case-Inventory},
  2025.
\newblock Compiled pursuant to E.O.\ 13960, the Advancing American AI Act, and
  OMB Memorandum M-25-21.

\bibitem[Reuel et~al.(2024)Reuel, Bucknall, Casper, Fist, Soder, Aarne,
  Hammond, Ibrahim, Chan, Wills, Anderljung, Garfinkel, Heim, Trask, Mukobi,
  Schaeffer, Baker, Hooker, Solaiman, Luccioni, Rajkumar, Mo{\"e}s, Ladish,
  Guha, Newman, Bengio, South, Pentland, Koyejo, Kochenderfer, and
  Trager]{reuelOpenProblemsTechnical2024}
Reuel, A., Bucknall, B., Casper, S., Fist, T., Soder, L., Aarne, O., Hammond,
  L., Ibrahim, L., Chan, A., Wills, P., Anderljung, M., Garfinkel, B., Heim,
  L., Trask, A., Mukobi, G., Schaeffer, R., Baker, M., Hooker, S., Solaiman,
  I., Luccioni, A.~S., Rajkumar, N., Mo{\"e}s, N., Ladish, J., Guha, N.,
  Newman, J., Bengio, Y., South, T., Pentland, A., Koyejo, S., Kochenderfer,
  M.~J., and Trager, R.
\newblock Open {{Problems}} in {{Technical AI Governance}}, July 2024.

\bibitem[Russell et~al.(2025)Russell, Karpinska, and
  Iyyer]{russellPeopleWhoFrequently2025}
Russell, J., Karpinska, M., and Iyyer, M.
\newblock People who frequently use {ChatGPT} for writing tasks are accurate
  and robust detectors of {AI}-generated text.
\newblock In \emph{Proceedings of the 63rd Annual Meeting of the Association
  for Computational Linguistics (Volume 1: Long Papers)}, pp.\  5342--5373,
  Vienna, Austria, 2025. Association for Computational Linguistics.
\newblock URL \url{https://aclanthology.org/2025.acl-long.267/}.

\bibitem[Sadasivan et~al.(2025)Sadasivan, Kumar, Balasubramanian, Wang, and
  Feizi]{sadasivan2025aigeneratedtextreliablydetected}
Sadasivan, V.~S., Kumar, A., Balasubramanian, S., Wang, W., and Feizi, S.
\newblock Can ai-generated text be reliably detected?, 2025.
\newblock URL \url{https://arxiv.org/abs/2303.11156}.

\bibitem[Sevilla et~al.(2022)Sevilla, Heim, Ho, Besiroglu, Hobbhahn, and
  Villalobos]{sevillaComputeTrendsThree2022}
Sevilla, J., Heim, L., Ho, A., Besiroglu, T., Hobbhahn, M., and Villalobos, P.
\newblock Compute {{Trends Across Three Eras}} of {{Machine Learning}}.
\newblock In \emph{2022 {{International Joint Conference}} on {{Neural
  Networks}} ({{IJCNN}})}, pp.\  1--8, July 2022.
\newblock \doi{10.1109/IJCNN55064.2022.9891914}.

\bibitem[Shah \& Levy(2026)Shah and Levy]{shahAccessJusticeAge2026}
Shah, A. and Levy, J.
\newblock Access to {{Justice}} in the {{Age}} of {{AI}}: {{Evidence}} from
  {{U}}.{{S}}. {{Federal Courts}}.
\newblock SSRN working paper, March 2026.
\newblock URL \url{https://ssrn.com/abstract=6766859}.

\bibitem[Sheehan(2023)]{sheehanChinaAIRegulations2023}
Sheehan, M.
\newblock China's {AI} regulations and how they get made.
\newblock Technical report, Carnegie Endowment for International Peace, July
  2023.
\newblock URL
  \url{https://carnegieendowment.org/research/2023/07/chinas-ai-regulations-and-how-they-get-made}.

\bibitem[Sutter(2025)]{sutterExportControlsChinaSemiconductors2025}
Sutter, K.~M.
\newblock U.s. export controls and china: Advanced semiconductors.
\newblock CRS Report R48642, Congressional Research Service, September 2025.
\newblock URL \url{https://www.congress.gov/crs-product/R48642}.
\newblock Updated September 19, 2025.

\bibitem[{Tobin-Miyaji}(2024)]{FederalAgenciesLargely}
{Tobin-Miyaji}, M.
\newblock Federal {{Agencies Largely Miss}} the {{Mark}} on {{Documenting AI
  Compliance Plans}} as {{Required}} by {{AI Executive Order}}.
\newblock EPIC, November 2024.
\newblock URL
  \url{https://epic.org/federal-agencies-largely-miss-the-mark-on-documenting-ai-compliance-plans-as-required-by-ai-executive-order/}.
\newblock Published November 21, 2024.

\bibitem[{U.S. Department of
  Defense}(2025)]{u.s.departmentofdefenseMilitarySecurityDevelopments2025}
{U.S. Department of Defense}.
\newblock Military and {{Security Developments Involving}} the {{People}}'s
  {{Republic}} of {{China}}.
\newblock Technical report, December 2025.

\bibitem[{U.S. Government Accountability
  Office}(2023)]{gaoAIImplementation2023}
{U.S. Government Accountability Office}.
\newblock {Artificial Intelligence: Agencies Have Begun Implementation but Need
  to Complete Key Requirements}.
\newblock Technical Report GAO-24-105980, {U.S. Government Accountability
  Office}, December 2023.
\newblock URL \url{https://www.gao.gov/products/gao-24-105980}.

\bibitem[{U.S. Government Accountability Office}(2025)]{gao2025generativeAI}
{U.S. Government Accountability Office}.
\newblock Artificial intelligence: Generative ai use and management at federal
  agencies.
\newblock Report to Congressional Requesters GAO-25-107653, U.S. Government
  Accountability Office, July 2025.
\newblock URL \url{https://www.gao.gov/assets/gao-25-107653.pdf}.

\bibitem[Wildeford(2026)]{CongressSuperintelligenceAI2026}
Wildeford, P.
\newblock Congress on {{Superintelligence}}.
\newblock The AI Policy Network (AIPN), June 2026.
\newblock URL \url{https://theaipn.org/issue/quotes/}.

\bibitem[Wu et~al.(2025)Wu, Yang, Zhan, Yuan, Chao, and
  Wong]{wuSurveyLLMGeneratedText2025}
Wu, J., Yang, S., Zhan, R., Yuan, Y., Chao, L.~S., and Wong, D.~F.
\newblock A {{Survey}} on {{LLM-Generated Text Detection}}: {{Necessity}},
  {{Methods}}, and {{Future Directions}}.
\newblock \emph{Computational Linguistics}, 51\penalty0 (1):\penalty0 275--338,
  March 2025.
\newblock ISSN 0891-2017.
\newblock \doi{10.1162/coli_a_00549}.

\bibitem[Zimmerman \& Mathur(2026)Zimmerman and
  Mathur]{Ombegov2024FederalAIUseCaseInventory2026}
Zimmerman, M. and Mathur, V.
\newblock Ombegov/2024-{{Federal-AI-Use-Case-Inventory}}.
\newblock Office of the Federal Chief Information Officer, April 2026.

\end{thebibliography}
\bibliographystyle{icml2026}

\clearpage
\appendix
\onecolumn

\section*{AI Usage Statement}
We used large language models to assist with data-collection scripting, figure generation, and prose drafting during the preparation of this paper. All substantive claims, analysis decisions, and interpretations are the authors' own.

\section{Related Work}
\label{app:related_work}

Our pilot study sits at the intersection of four threads of work: the technical AI governance agenda that motivates ``ecosystem monitoring'' as a research target, methods for detecting LLM-generated or LLM-assisted text, empirical studies that apply such methods to corpora of public-facing writing, and parallel efforts to measure government AI adoption through procurement, patents, public statements, and formal disclosure regimes.

\subsection{Technical AI Governance and Ecosystem Monitoring}

Our framing draws on the technical AI governance literature, in particular \citet{reuelOpenProblemsTechnical2024}, which identifies ``ecosystem monitoring''---collecting information about stakeholders, trends, and impacts in the AI ecosystem---as an underdeveloped capacity for AI governance.
Public-document AI detection, as we develop it here, is one concrete instantiation of usage-pattern trends directed specifically at the state.
It is complementary to other ecosystem-monitoring signals catalogued in that literature, including survey-based measures, direct usage monitoring~\citep{appelAnthropicEconomicIndex2025,misraMeasuringAIDiffusion2025}, reported attitudes~\citep{CongressSuperintelligenceAI2026} and more general indicators such as compute trends~\citep{sevillaComputeTrendsThree2022} and algorithmic progress measurements~\citep{hoAlgorithmicProgressLanguage2024}.

\subsection{AI-Text Detection}

This body of work develops detectors that classify whether a passage was written or substantially modified by an LLM.
Approaches include zero-shot statistical detectors that exploit log-probability curvature, such as DetectGPT~\citep{mitchellDetectGPTZeroShotMachineGenerated2023}; cross-model perplexity tests like Binoculars~\citep{hansSpottingLLMsBinoculars2024}; supervised classifiers trained on paired human/LLM corpora, including Pangram~\citep{emiTechnicalReportPangram2024}, which we use here; and watermarking schemes that aim to embed detectable signals at generation time~\citep{kirchenbauerWatermarkLargeLanguage2024}.
Survey work has catalogued the methods, datasets, and known failure modes of this literature~\citep{wuSurveyLLMGeneratedText2025}.
Previous work has found that detectors are vulnerable to paraphrasing, translation, and domain shift, and that calibration errors can display systematic biases in flagging particular author populations~\citep{liangGPTDetectorsAre2023}.
Our work necessarily inherits these limitations.

\subsection{Empirical Studies of LLM-Generated Text in the Wild}

A second thread applies these detectors at the corpus level.

\citet{liangMonitoringAIModifiedContent2024} introduce a maximum-likelihood approach for estimating the fraction of text in a large corpus that is substantially modified by an LLM, and apply it to AI-conference peer reviews, finding that 6.5--16.9\% of review text immediately after ChatGPT's release was likely LLM-modified.
\citet{liangQuantifyingLargeLanguage2025} take the same approach to scientific papers, documenting rapid uptake across disciplines.
\citet{ansariEchoesAutomationIncreasing2026} apply a panel of detectors to over 40{,}000 news articles and find a substantial rise in GenAI use, with local and college outlets adopting most aggressively, LLMs concentrated in introductions rather than conclusions, and stylistic homogenization especially in local media.
And \citet{shahAccessJusticeAge2026} document that the release of general-purpose LLMs has measurably altered the rate at which individuals self-represent in U.S.\ federal courts, providing complementary evidence that AI assistance is leaving fingerprints in formal public-record text streams.
Beyond academic and journalistic corpora, recent informal analyses have extended this kind of measurement to clinical trial registry entries, earnings-call prepared remarks, and federal court opinions from CourtListener, while reporting little signal in SEC filings~\citep{jsdAddingClinicalTrial2026}.
We see our work as filling a specific gap in this literature: applying the same family of techniques to a panel of government-authored documents from two of the political systems most directly relevant to AI governance.

\subsection{Monitoring Government AI Adoption and Attitudes}

A third area of the literature documents government AI use through means other than text detection, including procurement records, public statements, and formal disclosure regimes.

In the United States, EO 13960~\citep{PromotingUseTrustworthy2020} requires federal agencies to publish annual inventories of their non-classified AI use cases; a 2024 consolidated inventory now contains over 2{,}000 reported use cases~\citep{Ombegov2024FederalAIUseCaseInventory2026}.
Notably for the present work, independent audits have found substantial under-reporting and inconsistent categorization across agencies~\citep{FederalAgenciesLargely,gaoAIImplementation2023}.

\citet{Johnson_2025} use structured interviews to characterize AI procurement practices in US cities. \citet{cheungDeepSeekUsePRC2025} compile PLA procurement records explicitly calling for DeepSeek-based tools across the second half of 2025.
The U.S.\ Department of Defense's annual report on Chinese military power similarly identifies LLMs as a PLA capability area for coding, decision-support Q\&A, and synthetic content for influence operations~\citep{u.s.departmentofdefenseMilitarySecurityDevelopments2025}.

These signals differ from ours in their reliance on \emph{stated} adoption (procurement, self-reported inventories) rather than revealed behavior; consequently, they are more likely to be informative about which models are licensed or which use cases are formally acknowledged rather than which government bodies are routinely using LLMs in their day-to-day document production. Nevertheless, they have significant advantages in reliability and detail provided. Our detector-based signal is intended as a low-cost, externally reproducible complement to these efforts, not a substitute.

\section{Quality Check}
\label{app:data-quality-check}

To verify the integrity of our scraping and ingestion pipeline, we drew
a random sample of 5 documents per (source, year) pair---200
documents in total across our 10 sources and four years---and had
Claude Sonnet 4.6 (\texttt{claude-sonnet-4-6}) produce a structured
annotation for each via a forced \texttt{process\_document} tool call.
The annotation captures five fields: \texttt{is\_substantive}
(whether the body communicates real content about government
function, as opposed to a navigation page or empty stub),
\texttt{placeholder\_reason} (a short phrase explaining the
\texttt{false} case), \texttt{primary\_link} (the URL of an external
reference when the document body is essentially a pointer to it),
\texttt{parsing\_quality} (one of \texttt{clean}, \texttt{minor\_cruft},
\texttt{heavy\_cruft}, \texttt{failed}), \texttt{cruft\_notes}, and
\texttt{suspicious\_date}. 

We emphasize that this is a semi-automated spot-check of a small fraction of our dataset.
While we expect that there are multiple remaining errors, we believe that the high-level findings are nevertheless directionally correct.

\paragraph{Distributions.}
After the remediation passes described below, of the 200 sampled
documents 198 were marked substantive and 2 placeholder. Parsing
quality was 139 \texttt{clean}, 60 \texttt{minor\_cruft}, 1
\texttt{heavy\_cruft}, and 0 \texttt{failed}, with
\texttt{cruft\_notes} populated on 61 documents (the union of the two
\texttt{*\_cruft} buckets). Nineteen documents drew a
\texttt{suspicious\_date} flag.

\paragraph{Placeholders.}
The original sample surfaced 7 non-substantive cases. Five formed a
uniform structural pattern: a one-sentence introduction followed by a
hyperlink (rendered in the extracted text as the bare word
``here'') to an external PDF, hosted under
\texttt{bidenwhitehouse.archives.gov}, where the OSTP NSTC
announcement page is a content-free wrapper around a downloadable
PDF. The remaining two were CAC pages: a commentary item whose
article body failed to render at scrape time (the HTML contained only
navigation chrome and placeholder \texttt{<img>} tags), and a survey
page consisting only of a QR code and short instructions. We
expanded the check across the full corpus, not just the sample, using
a heuristic of ``text body $<$~600~bytes \emph{and} a PDF link present
in the cached HTML.'' This surfaced 29 stubs of the same shape: 21
OSTP, 4 MOST ethics-guideline announcements (each pointing to a
primary PDF plus a separate commentary PDF), 2 ministry notices on
\texttt{gov.cn}, and 2 CAC items where the PDF is loaded inside an
embedded \texttt{viewer.html?file=\ldots.pdf} \texttt{<iframe>}
rather than an \texttt{<a~href>} anchor. For 28 of these we replace
the stub with extracted PDF body text via \texttt{pdftotext}; the
remaining MOST entry (a 2025 ethics-guidelines announcement) is
unrecoverable because the source article page and both linked PDFs
have been removed from \texttt{www.most.gov.cn} since the original
scrape. The two CAC sample cases without a downloadable artifact
likewise remain flagged as unrecovered scrapes and are excluded from
downstream analysis; these are the 2 placeholder cases in the
post-remediation distribution.

\paragraph{Cruft.}
The original sample's 74 \texttt{minor\_cruft} cases were concentrated
in OSTP and \texttt{fedreg}, but the underlying issues were
different. OSTP cases were site-search bars and breadcrumb navigation
(e.g.\ ``To search this site, enter a search term / Search / Home /
OSTP / News \& Updates'') prepended to an otherwise intact article
body: the article-body CSS selectors were falling through to a
full-page text fallback because the actual content container is a
\texttt{<section class="body-content">} on the Biden archive and a
\texttt{<div class="entry-content">} with a
\texttt{<div class="wp-block-whitehouse-topper">} prepended on the
post-2025 \texttt{whitehouse.gov} block-theme template, neither of
which the original tag-specific \texttt{<div>} regex matched. We
tightened the selector to be tag-agnostic across both
\texttt{.body-content} and \texttt{.entry-content} and explicitly
strip topper, breadcrumb, and entry-header elements before text
extraction. \texttt{fedreg} cases were Cloudflare email-address
obfuscation: the \texttt{raw\_text\_url} endpoint passes through
Cloudflare, which replaces \texttt{user@dept.gov} addresses with
\texttt{<span data-cfemail="HEX">[email\ protected]</span>} elements
inside an \texttt{<a~href="/cdn-cgi/l/email-protection\#HEX">}
wrapper. Stripping anchor tags left the placeholder text in place.
We add a decoder for the obfuscation (the first byte of the hex
string is a single-byte XOR key, applied to each subsequent byte) and
substitute the recovered email before anchor stripping; this restored
313 contact addresses across the full \texttt{fedreg} corpus.
Reparsing both sources from the preserved raw HTML required no
re-fetching from the upstream sites. The 19 \texttt{minor\_cruft}
cases that remain in the post-fix \texttt{fedreg} sample are about
\emph{Federal Register} document conventions
([{[}Page \(N\){]}] page-break markers, [GRAPHIC] [TIFF OMITTED]
placeholders for plate inserts, billing codes), not parsing
artifacts; we preserve them as part of the published document text.

\paragraph{Suspicious dates.}
Of the 19 \texttt{suspicious\_date} flags, the model itself resolved
the majority as not genuine mismatches: archival paths often mark a
January-of-year-$N{+}1$ press release whose body discusses
year-$N$~events, which the prompt instructs to leave unflagged but
which appears to nudge a few cases through anyway. One real mismatch
surfaced---a CAC document under the 2024 archive path whose CMS
footer reads \texttt{publishdate:2025/07/10}, indicating either a
later edit or a backdated archival assignment. Two more flags were also false
positives: the model treated references to
the ``U.S.\ Department of War'' in two DARPA 2026 documents as
anomalous, on the grounds that no such department existed.

\section{Source Descriptions}
\label{app:source_descriptions}
We set a hard limit of 50,000 characters for each source document. In practice this means that we truncate 123 of the 386 Federal Register documents. All other collected documents fall under this limit. \Cref{tab:source_coverage} reports the document count for each (source, year) pair, and \cref{tab:ci_results} gives the corresponding mean AI fraction with bootstrap confidence intervals at four nested levels. The dataset cutoff date was 2026-04-25, meaning that we do not have complete coverage of 2026 for any source. The following subsections provide brief descriptions of each source.

\begin{table*}[t]
    \centering
    \resizebox{\textwidth}{!}{\begin{tabular}{llrrrrr}
\toprule
Source & Year & Mean & 50\% CI & 65\% CI & 80\% CI & 95\% CI \\
\midrule
CAC (CN) & 2021 & 0.01 & [0.00, 0.02] & [0.00, 0.02] & [0.00, 0.03] & [0.00, 0.04] \\
 & 2024 & 0.02 & \textbf{[0.01, 0.03]} & \textbf{[0.01, 0.03]} & \textbf{[0.01, 0.04]} & [0.00, 0.05] \\
 & 2025 & 0.07 & \textbf{[0.06, 0.09]} & \textbf{[0.05, 0.10]} & \textbf{[0.04, 0.11]} & \textbf{[0.03, 0.13]} \\
 & 2026 & 0.13 & \textbf{[0.11, 0.15]} & \textbf{[0.10, 0.16]} & \textbf{[0.09, 0.17]} & \textbf{[0.07, 0.19]} \\
\addlinespace
gov.cn (ministry) (CN) & 2021 & 0.00 & [0.00, 0.00] & [0.00, 0.00] & [0.00, 0.00] & [0.00, 0.00] \\
 & 2024 & 0.00 & [0.00, 0.00] & [0.00, 0.00] & [0.00, 0.00] & [0.00, 0.00] \\
 & 2025 & 0.00 & [0.00, 0.01] & [0.00, 0.01] & [0.00, 0.01] & [0.00, 0.01] \\
 & 2026 & 0.00 & [0.00, 0.01] & [0.00, 0.01] & [0.00, 0.01] & [0.00, 0.01] \\
\addlinespace
gov.cn (State Council) (CN) & 2021 & 0.00 & [0.00, 0.00] & [0.00, 0.00] & [0.00, 0.00] & [0.00, 0.00] \\
 & 2024 & 0.00 & [0.00, 0.01] & [0.00, 0.01] & [0.00, 0.01] & [0.00, 0.01] \\
 & 2025 & 0.01 & \textbf{[0.01, 0.01]} & \textbf{[0.01, 0.01]} & \textbf{[0.01, 0.02]} & [0.00, 0.02] \\
 & 2026 & 0.01 & \textbf{[0.01, 0.01]} & [0.00, 0.02] & [0.00, 0.02] & [0.00, 0.02] \\
\addlinespace
gov.cn (commentary) (CN) & 2021 & 0.00 & [0.00, 0.00] & [0.00, 0.00] & [0.00, 0.00] & [0.00, 0.00] \\
 & 2024 & 0.00 & [0.00, 0.00] & [0.00, 0.00] & [0.00, 0.00] & [0.00, 0.00] \\
 & 2025 & 0.02 & \textbf{[0.01, 0.02]} & \textbf{[0.01, 0.03]} & [0.00, 0.03] & [0.00, 0.04] \\
 & 2026 & 0.01 & \textbf{[0.01, 0.01]} & \textbf{[0.01, 0.02]} & [0.00, 0.02] & [0.00, 0.02] \\
\addlinespace
MOST (CN) & 2021 & 0.00 & [0.00, 0.00] & [0.00, 0.00] & [0.00, 0.00] & [0.00, 0.00] \\
 & 2024 & 0.03 & \textbf{[0.02, 0.04]} & \textbf{[0.01, 0.04]} & \textbf{[0.01, 0.05]} & [0.00, 0.06] \\
 & 2025 & 0.04 & \textbf{[0.03, 0.06]} & \textbf{[0.03, 0.06]} & \textbf{[0.02, 0.07]} & \textbf{[0.01, 0.08]} \\
 & 2026 & 0.11 & \textbf{[0.07, 0.14]} & \textbf{[0.06, 0.16]} & \textbf{[0.05, 0.17]} & \textbf{[0.02, 0.21]} \\
\addlinespace
PLA Daily (CN) & 2021 & 0.00 & [0.00, 0.00] & [0.00, 0.00] & [0.00, 0.00] & [0.00, 0.00] \\
 & 2024 & 0.01 & [0.00, 0.01] & [0.00, 0.01] & [0.00, 0.01] & [0.00, 0.02] \\
 & 2025 & 0.16 & \textbf{[0.14, 0.18]} & \textbf{[0.13, 0.19]} & \textbf{[0.12, 0.20]} & \textbf{[0.10, 0.22]} \\
 & 2026 & 0.14 & \textbf{[0.12, 0.16]} & \textbf{[0.12, 0.17]} & \textbf{[0.11, 0.18]} & \textbf{[0.09, 0.21]} \\
\addlinespace
DARPA news (US) & 2021 & 0.00 & [0.00, 0.00] & [0.00, 0.00] & [0.00, 0.00] & [0.00, 0.00] \\
 & 2024 & 0.00 & [0.00, 0.00] & [0.00, 0.00] & [0.00, 0.00] & [0.00, 0.00] \\
 & 2025 & 0.04 & \textbf{[0.02, 0.06]} & \textbf{[0.02, 0.06]} & [0.00, 0.09] & [0.00, 0.11] \\
 & 2026 & 0.08 & \textbf{[0.05, 0.12]} & \textbf{[0.03, 0.14]} & [0.00, 0.17] & [0.00, 0.20] \\
\addlinespace
Federal Register (US) & 2021 & 0.00 & [0.00, 0.00] & [0.00, 0.00] & [0.00, 0.00] & [0.00, 0.00] \\
 & 2024 & 0.00 & [0.00, 0.00] & [0.00, 0.00] & [0.00, 0.00] & [0.00, 0.00] \\
 & 2025 & 0.00 & [0.00, 0.01] & [0.00, 0.01] & [0.00, 0.01] & [0.00, 0.01] \\
 & 2026 & 0.01 & \textbf{[0.01, 0.02]} & \textbf{[0.01, 0.02]} & [0.00, 0.03] & [0.00, 0.03] \\
\addlinespace
Military Review (US) & 2021 & 0.00 & [0.00, 0.00] & [0.00, 0.00] & [0.00, 0.00] & [0.00, 0.00] \\
 & 2024 & 0.01 & \textbf{[0.01, 0.02]} & [0.00, 0.02] & [0.00, 0.03] & [0.00, 0.04] \\
 & 2025 & 0.04 & \textbf{[0.03, 0.05]} & \textbf{[0.03, 0.05]} & \textbf{[0.03, 0.06]} & \textbf{[0.02, 0.07]} \\
 & 2026 & 0.10 & \textbf{[0.08, 0.13]} & \textbf{[0.07, 0.14]} & \textbf{[0.06, 0.15]} & \textbf{[0.04, 0.17]} \\
\addlinespace
OSTP (US) & 2021 & 0.00 & [0.00, 0.00] & [0.00, 0.00] & [0.00, 0.00] & [0.00, 0.00] \\
 & 2024 & 0.00 & [0.00, 0.00] & [0.00, 0.00] & [0.00, 0.00] & [0.00, 0.00] \\
 & 2025 & 0.00 & [0.00, 0.00] & [0.00, 0.00] & [0.00, 0.00] & [0.00, 0.00] \\
 & 2026 & 0.00 & [0.00, 0.00] & [0.00, 0.00] & [0.00, 0.00] & [0.00, 0.00] \\
\bottomrule
\end{tabular}
}
    \caption{Per-source mean AI fraction by year, with 50\%, 65\%, 80\%, and 95\% bootstrap confidence intervals. Intervals which exclude 0 are bolded. Values are the mean Pangram \texttt{fraction\_ai} $\in [0,1]$; intervals come from a standard bootstrap over the documents in each (source, year) cell.}
    \label{tab:ci_results}
\end{table*}

\begin{CJK}{UTF8}{gbsn}

\paragraph{CAC.}
The Cyberspace Administration of China (国家互联网信息办公室, CAC) is the
central regulator of internet content, data, and algorithm governance
in the People's Republic of China. Its public website
(\url{www.cac.gov.cn}) publishes binding rules and normative documents
(政策法规, ``policy and regulations''), official news items (网信要闻, ``cyberspace
headlines'', and 网信发布, ``cyberspace announcements''),
interpretive commentary (理响中国, ``Theoretical Voice China''), and a
set of topical sub-channels covering content governance, the recurring
清朗 (``clear and bright'') enforcement campaigns,
network security, data governance, informatization, international
exchange, and miscellaneous work. We pool eleven of these channels
into a single candidate set per year. We deliberately exclude
时政要闻 (top-level political news --- typically Xi
speeches, which appear verbatim across every ministry site),
规范性文件 (a thirty-item set of documents
already fully contained in the policy and regulations channel), and
互动服务 (a citizen-complaint UI with no textual
content).

\paragraph{Ministry-issued documents (gov.cn).}
This corpus draws from the 部门文件 bucket of the Chinese
Government Web Portal policy library
(\url{www.gov.cn/zcwjk/policyDocumentLibrary}), which contains
ministry-issued documents republished to gov.cn. The collection is restricted to AI-related
material via an OR-union over the keyword set \{通用人工智能,
大语言模型, 大模型, 生成式人工智能, 人工智能\} (roughly, ``artificial
general intelligence'', ``large language model'', ``large model'' /
``foundation model'', ``generative artificial intelligence'', and
``artificial intelligence'' respectively). Because we collect MOST
(科技部, the Ministry of Science and Technology) and CAC (国家互联网信息办公室,
the Cyberspace Administration of China) separately as first-class
sources, we filter out single-issuer documents from these two
ministries at discovery time to avoid double counting.

\paragraph{State Council policy documents (gov.cn).}
The Chinese Government Web Portal (中国政府网, \url{www.gov.cn}) is the
official aggregator for State Council and ministry-level documents.
Its policy library (政策文件库) splits results into four buckets;
this corpus draws from the \texttt{gongwen} bucket, which contains the
State Council's own normative instruments (the 国发, 国办发, 国令,
国函, and 国办函 document series --- roughly, full State Council
issuances, general-office issuances, State Council orders, and the
two corresponding letter series). The collection is restricted to
AI-related documents via an OR-union over the keyword set
\{通用人工智能, 大语言模型, 大模型, 生成式人工智能, 人工智能\}
(roughly, ``artificial general intelligence'', ``large language
model'', ``large model'' / ``foundation model'', ``generative
artificial intelligence'', and ``artificial intelligence''
respectively).

\paragraph{State Council policy commentary (gov.cn).}
This corpus draws from a commentary bucket of the Chinese
Government Web Portal policy library
(\url{www.gov.cn/zcwjk/policyDocumentLibrary}). The bucket consists of
政策解读 (\textit{zhengce jiedu}, ``policy interpretation'') ---
official interpretive commentary essays paired with each policy
document --- together with a small number of news pieces hung off
individual policies. As with the other gov.cn buckets, the collection
is restricted to AI-related material via an OR-union over the keyword
set \{通用人工智能, 大语言模型, 大模型, 生成式人工智能, 人工智能\}
(roughly, ``artificial general intelligence'', ``large language
model'', ``large model'' / ``foundation model'', ``generative
artificial intelligence'', and ``artificial intelligence''
respectively). Commentary essays are useful as a complement to the
policy texts themselves: they tend to be more discursive and candid.

\paragraph{MOST.}
The Ministry of Science and Technology of the People's Republic of
China (中华人民共和国科学技术部, MOST) is the ministry that administers
national science and technology policy and funding; since the March
2023 institutional reform it also serves as the administrative office
of the Party's Central Science and Technology Commission. We collect its work
briefings (科技部工作, \textit{kjbgz}) at \url{www.most.gov.cn/kjbgz/},
which serve as a running ministry-level record of policy roll-outs,
program announcements, and field activities. The MOST live site
only retains content from late 2023 onward, so we draw the entire
corpus through the Internet Archive's Wayback Machine.

\paragraph{PLA Daily.}
\textit{PLA Daily} (解放军报) is the official
newspaper of the People's Liberation Army. Our corpus targets the
paper's commentary content (评论 ``commentary'', 时评 ``current-affairs
commentary'', and 要论 ``key articles''), which carries authoritative
interpretation of military policy and is read as an indicator of the
official line on doctrine, strategy, and political work. We 
collect exclusively through the Internet Archive's Wayback Machine,
since the live editions are not stably retained.

\paragraph{DARPA.}
The Defense Advanced Research Projects Agency (DARPA) is the U.S.
Department of Defense agency responsible for high-risk, high-reward
research. Its public news section (\url{www.darpa.mil/news})
publishes program announcements, research milestones, solicitations
of interest, and personnel notes attributed to specific technical
offices (e.g.\ I2O, MTO, STO, BTO). After DARPA's late-2024 site
redesign, news article URLs follow the slug pattern
\texttt{/news/YYYY/<slug>}, and prior articles have been re-published
under the same scheme.

\paragraph{Federal Register.}
The Federal Register (\url{www.federalregister.gov}) is the official
daily journal of the U.S.\ federal government, publishing rules,
proposed rules, notices, and presidential documents. We collect
AI-related Federal Register documents via the public REST API at
\texttt{/api/v1/documents.json}, OR-unioning over the term set
\{``artificial intelligence'', ``machine learning'', ``generative
AI'', ``large language model'', ``foundation model''\}. Documents are
restricted to a target year by API-side date filtering on
\texttt{publication\_date}.

\paragraph{Military Review.}
\textit{Military Review} is the bi-monthly professional journal of the
U.S.\ Army University Press (\url{www.armyupress.army.mil}). The
journal publishes essays on doctrine, military history, leadership,
and emerging technology, written largely by serving officers and
defense-policy researchers. Articles are organized by year and by
issue (six bi-monthly issues per year:
January--February, March--April, May--June, July--August,
September--October, and November--December). The publication carries
no per-article timestamps in the page chrome --- dates are
issue-level only.

\paragraph{OSTP.}
The White House Office of Science and Technology Policy (OSTP) is the
executive office responsible for advising the President on science
and technology policy. Its public news output is split across two
hosts in our window of interest: the Biden-era archive at
\url{bidenwhitehouse.archives.gov/ostp/news-updates/} (covering
January 2021 -- January 2025), and the Trump-2 site at
\url{www.whitehouse.gov/ostp/news/} (covering January 2025 onward).

\end{CJK}

\section{Per-Source Results}
\label{app:per_source_results}

\begin{figure}[htbp]
    \centering
    \includegraphics[width=0.48\textwidth]{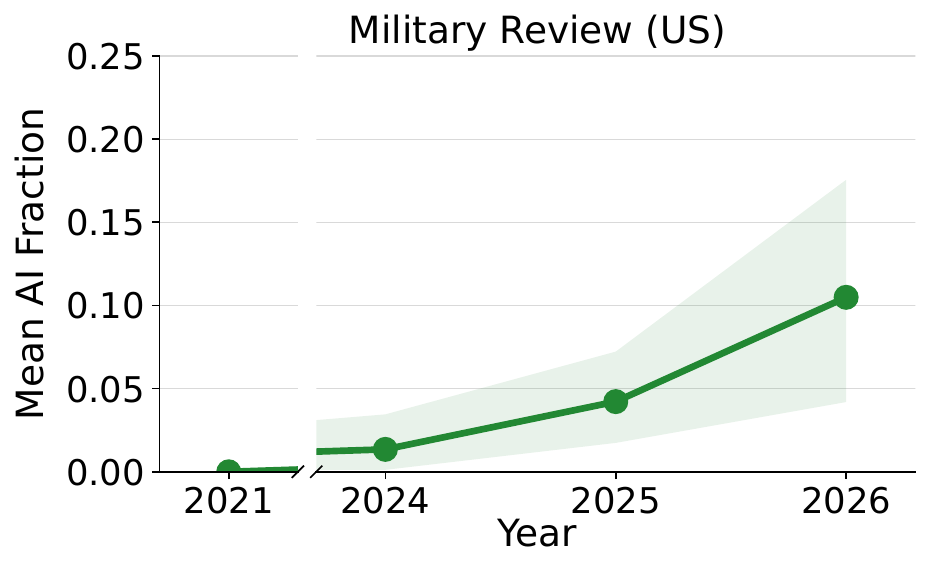}\hfill
    \includegraphics[width=0.48\textwidth]{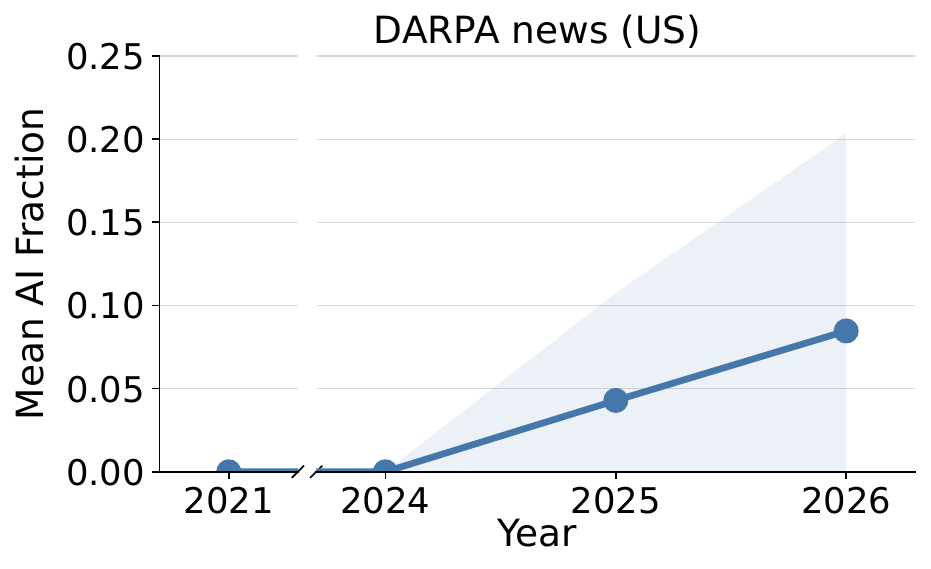}

    \medskip
    \includegraphics[width=0.48\textwidth]{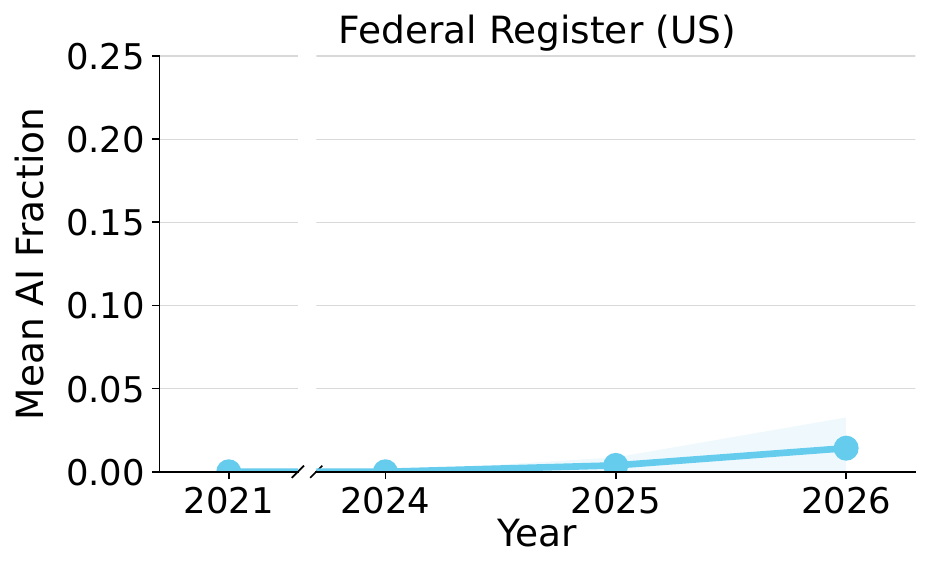}\hfill
    \includegraphics[width=0.48\textwidth]{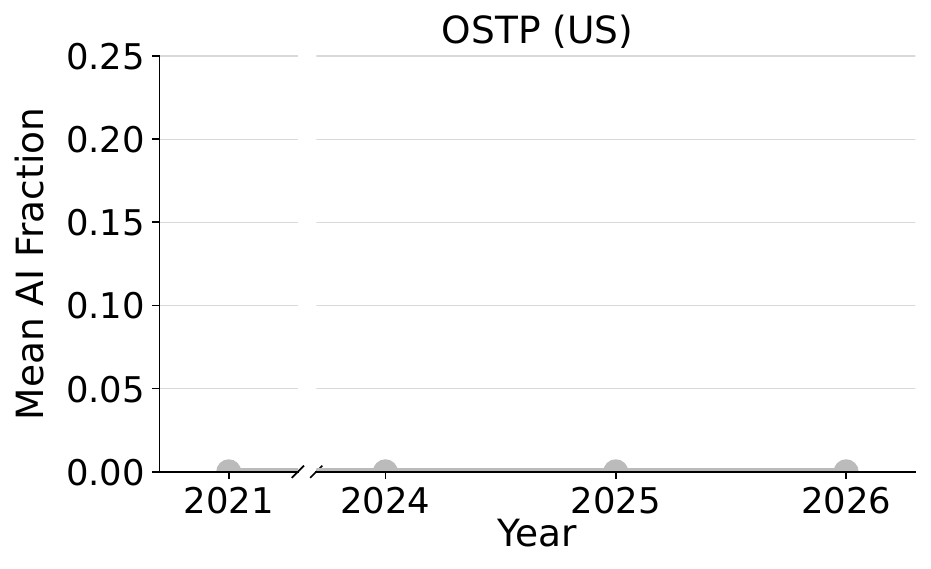}
    \caption{Per-source mean AI fraction over time for the four U.S.\ sources
    (\textit{Military Review}, DARPA news, Federal Register, OSTP), with 95\%
    bootstrap confidence intervals. All panels share a common $y$-axis.}
    \label{fig:by_source_plots}
\end{figure}

\begin{figure}[htbp]
    \centering
    \includegraphics[width=0.48\textwidth]{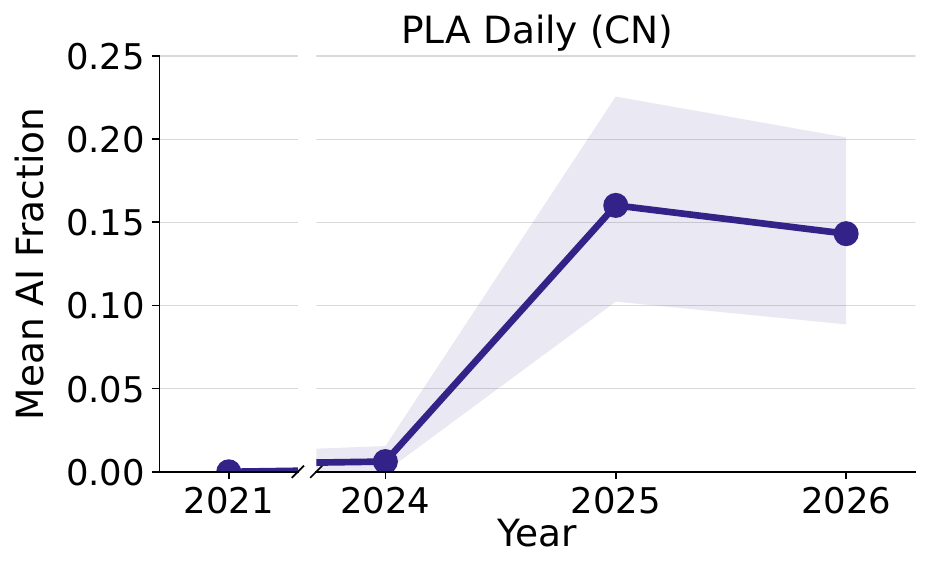}\hfill
    \includegraphics[width=0.48\textwidth]{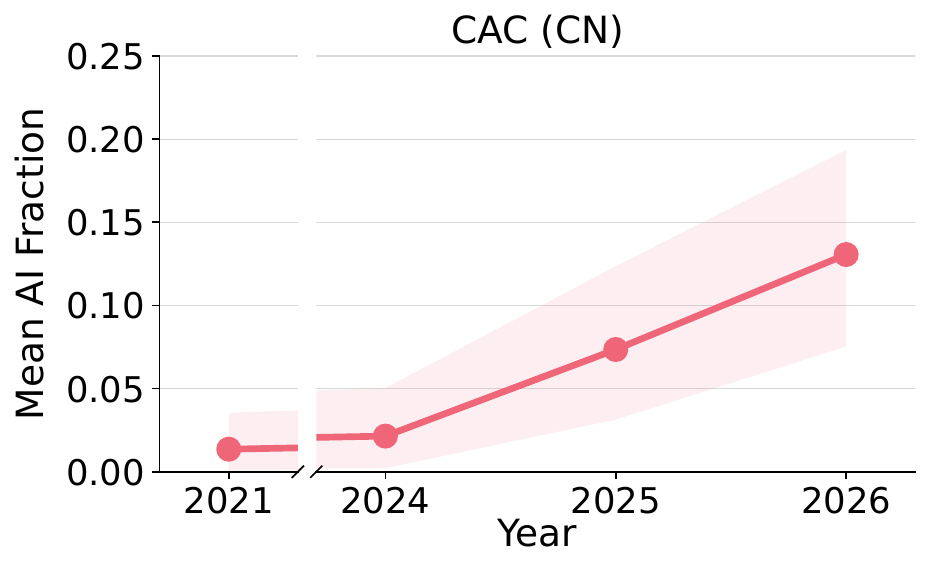}

    \medskip
    \includegraphics[width=0.48\textwidth]{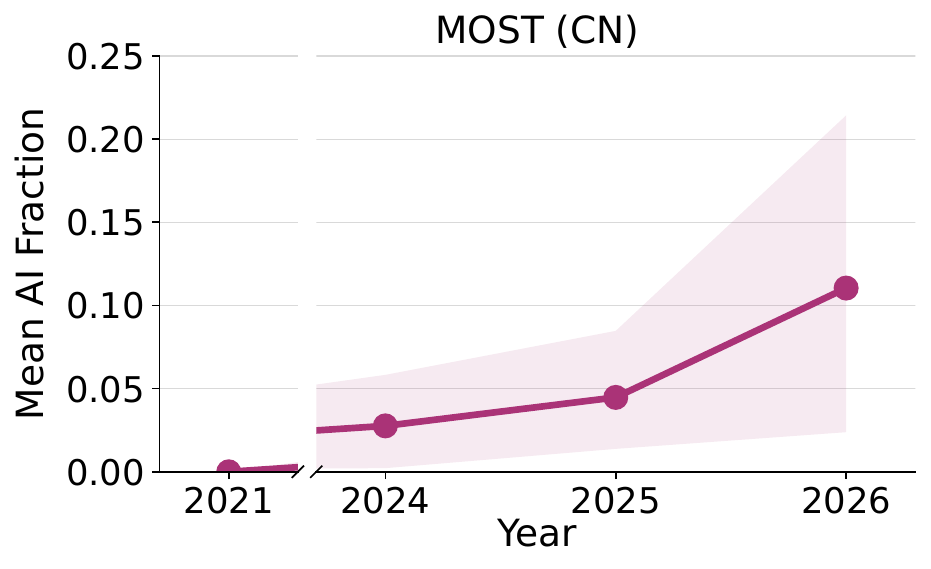}\hfill
    \includegraphics[width=0.48\textwidth]{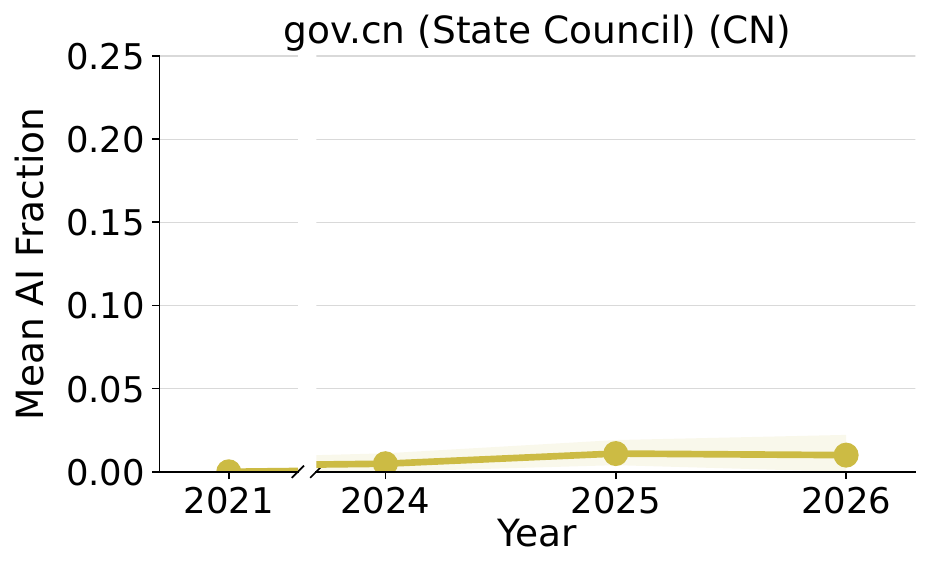}

    \medskip
    \includegraphics[width=0.48\textwidth]{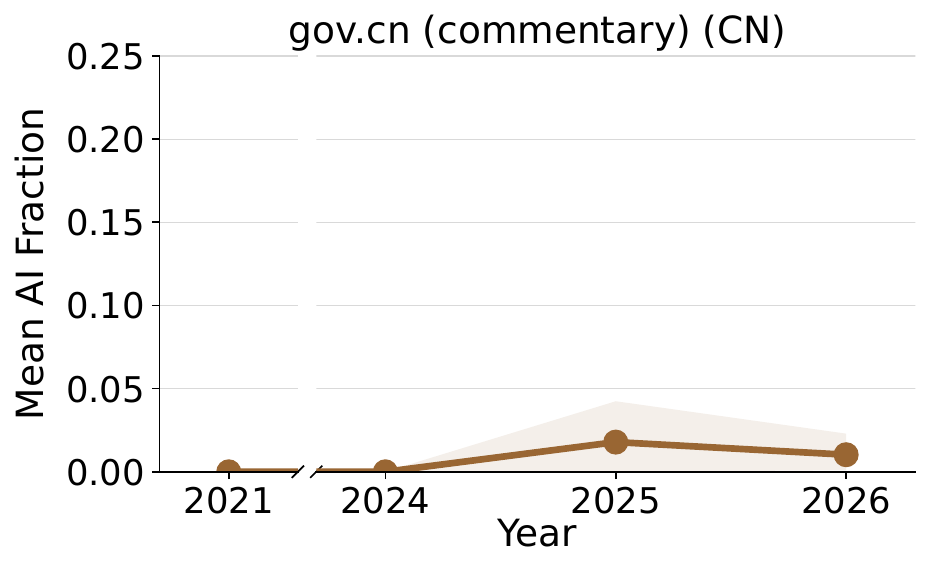}\hfill
    \includegraphics[width=0.48\textwidth]{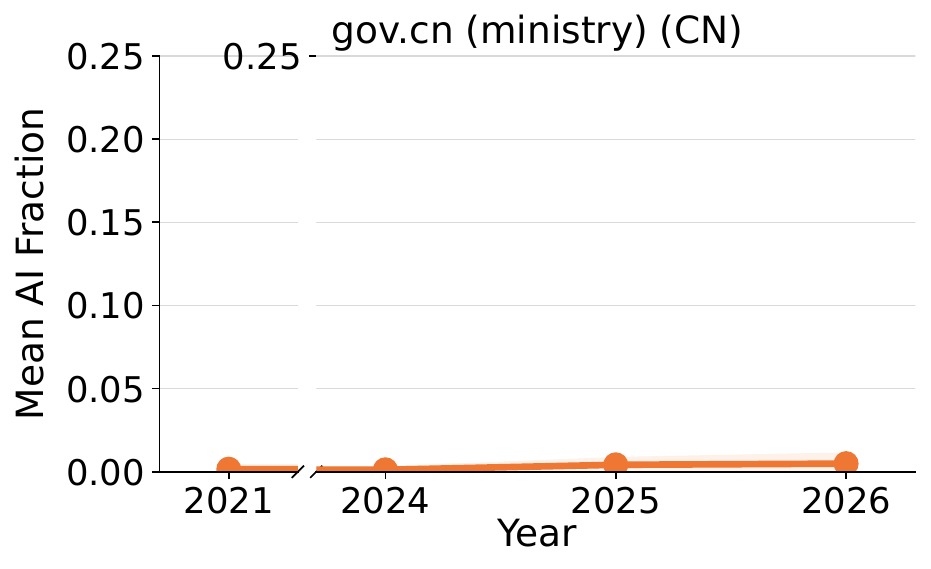}
    \caption{Per-source mean AI fraction over time for the six PRC sources
    (\textit{PLA Daily}, CAC, MOST, and the three gov.cn streams), with 95\%
    bootstrap confidence intervals. All panels share the same $y$-axis scale as
    \cref{fig:by_source_plots}.}
    \label{fig:by_source_prc}
\end{figure}

\Cref{fig:by_source_plots,fig:by_source_prc} show the per-source mean AI
fraction over time---with 95\% bootstrap confidence intervals---for the U.S.\
and PRC sources respectively, ordered within each panel group by descending
2026 score. All panels share the same $y$-axis scale to facilitate comparison
across sources. See \cref{app:source_descriptions} for descriptions of each
source. The numeric counterpart---per-source mean AI fraction by year with
50\%, 65\%, 80\%, and 95\% bootstrap confidence intervals---is given in
\cref{tab:ci_results} (\cref{app:source_descriptions}).

\section{Random Examples}
\label{app:random_examples}

For each source we show one randomly drawn document with \texttt{fraction\_ai} $<$~0.1 and up to 2 randomly drawn documents with \texttt{fraction\_ai} $>$~0.5 (both conditional such documents existing). Bodies are truncated at 2000 characters; Chinese-language documents are followed by an English translation produced by Claude Sonnet 4.6. (Note that Pangram detection was run on the original Chinese-language documents, not the translated versions.)

\subsection*{CAC (CN)}
\paragraph{Low-AI example (2026, \texttt{fraction\_ai}~$=$~0.00).}

\begin{quote}\small
\begin{CJK}{UTF8}{gbsn}
2025中国正能量网络精品征集展播活动结果出炉，现对拟入选作品进行公示，公示时间为2月10日至2月14日，公示期间欢迎社会各界监督评议，有关意见可发送至邮箱zgznl2025@chinanews.com.cn，我们将及时受理、认真核查，公示期满后将适时对外发布最终结果，并在线上进行集中展播。

中央网信办

2026年2月10日
\end{CJK}
\end{quote}
\noindent\textit{English translation.}

\begin{quote}\small
The results of the 2025 China Positive Energy Online Featured Works Collection and Exhibition Activity have been announced. The works proposed for selection are hereby published for public notice. The public notice period runs from February 10 to February 14. During the notice period, supervision and comments from all sectors of society are welcome; relevant opinions may be sent to the email address zgznl2025@chinanews.com.cn. We will handle submissions promptly and conduct thorough verification. After the public notice period expires, the final results will be released to the public in due course, and a concentrated online exhibition and broadcast will be held.

Cyberspace Administration of China

February 10, 2026
\end{quote}

\paragraph{High-AI example (2026, \texttt{fraction\_ai}~$=$~1.00).}

\begin{quote}\small
\begin{CJK}{UTF8}{gbsn}
近日，国家网信办等11部门联合印发了《关于提升境外人员入境数字化服务便利性的实施意见》（以下简称《实施意见》）。《实施意见》旨在建立互联互通、包容普惠、标准互认的数字化服务体系，打造更加国际化、便利化的数字化服务环境，从而以数字开放赋能高水平对外开放。

一、提升入境数字化服务是扩大高水平开放的必然要求

党的二十届四中全会明确提出要实现“高水平对外开放体制机制更加健全”。随着我国成为全球主要投资目的地、旅游大国和人才集聚地，每年有大量境外人员入境从事商务、旅游、学习、工作等活动。然而，长期以来，境外人员入境使用数字化服务仍面临一些堵点与痛点，如支付方式不互通、多语种服务缺失、平台使用门槛高等。这些问题不仅影响了境外人员入境工作生活体验，也在一定程度上制约了我国开展更高频次、更广范围的国际人员往来与经贸合作。

《实施意见》的出台，正是直面上述问题，以系统化思维推动数字赋能入境服务，从便捷电信业务办理、提升数字支付便利性到拓展文旅线上服务渠道，构建起覆盖入、购、游、行、住、医、学、创等的全场景数字化服务链条。这一政策的实施，将极大提升我国在全球化人才竞争、国际旅游市场和国际经贸合作中的软实力与吸引力，推动制度型开放迈出坚实步伐。

二、《实施意见》彰显四大政策亮点，系统构建入境数字化服务新生态

一是坚持系统集成，推动服务由“碎片化”向“一体化”跃升。

传统政务服务与数字平台建设往往存在部门分割、地域差异问题，导致境外人员需面对多个系统、重复注册、标准不一的困扰。《实施意见》的一个突出亮点在于强调系统观念，推动跨部门、跨层级、跨地区的服务整合与数据共享。例如，在“夯实数字化基础服务”部分，《实施意见》明确提出建设数字化综合服务平台，整合新闻资讯、语言翻译、政策指引、生活服务、消费指南等内容，并接入常用移动应用程序和网站。这旨在打造面向境外人员的一站式数字化服务门户，改变以往“散、乱、孤”的服务状态。这种顶层设计与底层打通相结合的模式，体现了以用户为中心、以场景为驱动的现代治理理念。

二是聚焦场景驱动，实现从“有”到“优”的精准赋能。

《实施意见》并未停留在原则性指导，而是深入境外人员入境工作生活的具体场景，提出了精准化、精细化的解决方案，推动数字化服务从可用向好用、易用、愿用转变。在完善数字化支付服务方面，《实施意见》不仅提出支持更多境外电子钱包在境内使用，还推动扩大免押金等优惠政策适用范围。在优化数字化旅游服务中，从线上预约、多语种推送、线路规划到国际银行卡购票、住宿线上预登记，基本覆盖入境旅游的全部流程。对于丰富数字化公共服务，《实施意见》更就中文数字学习资源的供给、企业开办的线上便利化等作出部署。这种基于真实需求、解决实际问题的场景化安排，确保了政策举措的针对性与实效性。

三是强化技术赋能，突出智能引领与标准先行。

《实施意见》高度重视前沿技术在提升服务体验、突破语言障碍、优化服务流程中的核心驱动作用，体现出鲜明的技术赋能导向。《实施意见》多次提及人工智能等技术应用，如“利用人工智能等技术提高外文翻译的即时性和准确性”“深化人工智能、虚拟现实、大数据等技术在消费领域应用”“积极探索‘人工智能+政务’应用”。这表明，政策鼓励运用AI大模型、虚拟现实等创新手段，打造更智能、更沉浸的服务体验。同时，《实施意见》也注重标准与规则的建设，如“编制专有名词翻译指南”“推动境外人员入住流程规范化、透明化、标准化”“积极参与数据跨境流动国际规则和标准的制定”。通过技术突破与标准共建双轮驱动，既解决了当前服务的燃眉之急，也为构建长期、稳定、与国际接轨的数字化服务生态奠定了基石。

四是统筹发展和安全，构筑开放可信的数字环境。

高水平开放必须建立在高水平安全的基础之上。境外人员数字化服务涉及大量跨境数据流动与个人信息处理，对网络安全、数据安全和个人信息保护提出了更高要求。《实施意见》专设“加强网络和数据安全保障”部分，展现出统筹发展和安全的清醒认识和战略定力。在网络安全方面，强调对跨境支付、在线预约等高频场景的重点防护，并深化国际合作，打击跨境网络安全犯罪。在数据安全与个人信息保护方面，要求加强数据分类分级和全流程管理，特别防范敏感信息泄露，并积极参与国际规则制定，强化数据出境安全监管。这一系列举措，旨在构建一个让境外人员敢用、放心用的可信数字环境，只有在安全底线牢固的前提下，数字化服务的便利性才能真正转化为我国对外开放的竞争优势。

三、《实施意见》为高水平开放注入数字动能

《实施意见》的出台，是我国顺应数字时代发展潮流、推动高水平对外开放的又一务实举措。作为我国首份系统针对境外人员入境全流程数字化服务的专项政策文件，《实施意见》围绕数字化基础服务、支付、旅游、公共服务及网络安全等关键环节，系统打造更加国际化、便利化、普惠化的数字化服务环境，对我国以数\,\ldots
\end{CJK}
\end{quote}
\noindent\textit{English translation.}

\begin{quote}\small
Recently, the Cyberspace Administration of China and 10 other departments jointly issued the "Implementation Opinions on Improving the Convenience of Digital Services for Inbound Foreign Nationals" (hereinafter referred to as the "Implementation Opinions"). The Implementation Opinions aim to establish a digital service system that is interconnected, inclusive, and mutually recognizes standards, and to create a more internationalized and convenient digital service environment, thereby using digital openness to empower a high-standard opening-up.

**I. Improving Digital Services for Inbound Foreign Nationals Is an Inevitable Requirement for Expanding High-Standard Opening-Up**

The Fourth Plenary Session of the 20th Central Committee of the Communist Party of China explicitly proposed achieving "a more sound institutional mechanism for high-standard opening-up to the outside world." As China has become a major global investment destination, a major tourism country, and a hub for talent, a large number of foreign nationals enter China each year to engage in business, tourism, study, work, and other activities. However, for a long time, foreign nationals entering China have still faced a number of bottlenecks and pain points when using digital services, such as incompatible payment methods, lack of multilingual services, and high barriers to using platforms. These issues have not only affected the work and life experience of inbound foreign nationals, but have also, to a certain extent, constrained China's ability to conduct higher-frequency and broader-ranging international personnel exchanges and economic and trade cooperation.

The issuance of the Implementation Opinions directly confronts the above-mentioned problems, using systematic thinking to drive digital empowerment of inbound services --- from facilitating telecommunications service handling and improving the convenience of digital payments to expanding online service channels for culture and tourism --- thereby\,\ldots
\end{quote}

\paragraph{High-AI example (2026, \texttt{fraction\_ai}~$=$~0.79).}

\begin{quote}\small
\begin{CJK}{UTF8}{gbsn}
人工智能的演进已出现微妙而深刻的转折，它从解决明确任务的“工具”，转向能够模拟自然人人格、思维模式和沟通风格的“互动者”。以高级语言模型和情感计算为核心技术的人工智能拟人化互动服务，正悄然进入老年陪伴、心理健康、情感照护等社会生活的敏感领域。拟人化人工智能在提供慰藉、缓解孤独的同时，也潜藏着诱导情感依赖、扭曲社交认知乃至危及用户身心健康的巨大风险。产生这种“双刃剑”效应的根源在于，人工智能拟人化互动服务（以下简称拟人化互动服务）超越了单纯的信息处理范畴，已深度触及人类最本质的情感与关系构建需求。

在此背景下，国家互联网信息办公室等五部门联合公布的《人工智能拟人化互动服务管理暂行办法》（以下简称《办法》），是一次极具前瞻性和科学性的立法响应。它不仅是对既有网络安全、数据保护、算法治理法律框架的补强，更是我国针对“人机情感交互”这一新型社会关系进行的专门化立法尝试。可以看出，我国人工智能立法正迈向主动化、体系化治理，呈现出以“负责任创新”为价值导向、以“系统化规制”为方法指引的核心脉络。

一、从工具化的监管对象到系统性的创新保障

《办法》关注到拟人化人工智能的风险并非仅源于最终生成的输出结果，而是根植于其作为“社会化智能体”的整个运行过程。因此，基于系统论理念，应将拟人化互动服务视为一个由数据、算法、算力、商业目标与用户反馈共同构成的、动态演化的复杂系统。在此视角下，《办法》的规制逻辑呈现出清晰的“全生命周期治理”特征，其核心目标在于对拟人化互动服务进行价值校准与行为纠偏。

首先是前置设定价值锚点，从源头矫正发展目标。

拟人化人工智能的许多潜在社会风险源于研发设计阶段将用户留存时长、情感粘性等作为优化指标，《办法》明确禁止“将替代社会交往、控制用户心理、诱导沉迷依赖等作为服务目标”，直指拟人化人工智能部署应用的重点问题。立法在系统的初始条件中就为技术植入了“以人为本”的理念，确保技术创新始终以增进人类福祉为价值导向，这种思路与人工智能“负责任创新”理念高度契合，彰显了从末端治理向源头治理的人工智能治理范式转变。

其次是动态干预运行过程，构建“感知---响应”闭环。

人工智能拟人化互动是一个持续的、情境化的过程，静态内容过滤难以应对动态交互产生的情感依赖与心理风险。对此，《办法》围绕“风险识别、及时干预”建构义务体系，要求拟人化互动服务提供者及时识别用户的极端情绪，在此基础上及时采取包括安抚、鼓励、提供相应渠道乃至联络紧急联系人等在内的分级响应措施。该举措实质上是在立法层面为人工智能系统增设了“社会安全员”功能，将其从一个纯粹的“交互者”部分引导为负有关照义务的“守护者”。从系统论角度看，这一制度设计旨在通过动态识别系统输出与预设目标之间的偏差并及时反馈、调整系统行为，从而维持人工智能系统整体的稳定与安全。

最后是全要素协同规制，贯穿数据、算法与交互界面。

人工智能系统的输出风险往往是数据偏见、算法黑箱与诱导性交互设计等共同发挥作用的结果，《办法》的制度设计完整覆盖了技术全链条。从训练数据质量与合法性要求，到算法机制机理审核，再到生成内容的标识与用户便捷退出机制，这种“组合拳”式的治理思路表明，必须通过多要素协同治理才能系统性降低人工智能的社会化风险。特别是要求服务提供者“采取有效措施提示用户正在与人工智能服务而非自然人进行互动”，它不只是一项纯粹的技术操作，更是在社会层面明晰“人机边界”的认知基石，旨在防止情感混淆、捍卫人类在人机互动中的主体性地位。

二、关键条款分类及差异化技术合规路径

《办法》如何在技术实践中落地生根，是检验其治理效能的关键。法律的包容性与工程实践对确定性指标的渴求，二者之间存在一定程度的张力。为了调和这一张力，可将《办法》的条款大致归为三类，其合规转化路径应当有所区别。

第一类是具备良好标准化基础的“行为指令”条款。

这类条款规定了明确、具体的行为模式，最易转化为具有技术操作性的规范。例如，《办法》关于“提示”和“动态提醒”的要求，可以通过“人机交互设计指南”来细化落实，明确提示信息的视觉显著性标准（如尺寸、颜色、位置）、持续时间及触发频率等指标。同样，基于对“用户极端情景识别”的评估要求，行业可共同探索关于状态识别准确率、风险语料特征库、预警阈值设置（如连续使用时长、负面情感词汇密度）等方面的评估标准。诸如此类规则的标准化延展，不仅能为企业提供清晰的行为指引，也能为监管提供相对客观的审查尺度。

第二类是依赖价值判断与过程管控的“边界禁止”条款。

这类条款如禁止生成鼓励、美化、暗示损害身心健康的内容和过度迎合用户等，其内涵具有情境弹性，通过负面清单难以穷尽列举所有的禁止情形，也容易被规避。这就要求合规建设时需证明服务的算法设计理念、内容过滤策略、风险干预模型等是围绕避免伤害、尊重自主等要求构建的，并且具备持续审计\,\ldots
\end{CJK}
\end{quote}
\noindent\textit{English translation.}

\begin{quote}\small
The evolution of artificial intelligence has reached a subtle yet profound turning point: it has shifted from being a "tool" for solving clearly defined tasks to an "interactor" capable of simulating the personality, thought patterns, and communication styles of real human beings. AI anthropomorphic interaction services---whose core technologies are advanced language models and affective computing---are quietly entering sensitive domains of social life such as elderly companionship, mental health, and emotional care. While anthropomorphic AI provides comfort and alleviates loneliness, it also harbors enormous risks of inducing emotional dependence, distorting social cognition, and even endangering users' physical and mental wellbeing. The root cause of this "double-edged sword" effect is that AI anthropomorphic interaction services (hereinafter referred to as anthropomorphic interaction services) transcend the realm of pure information processing and have penetrated deeply into the most fundamental human needs for emotion and relationship formation.

Against this backdrop, the Interim Measures for the Administration of Artificial Intelligence Anthropomorphic Interaction Services (hereinafter referred to as the Measures), jointly promulgated by the Cyberspace Administration of China and four other departments, represents a legislative response of exceptional foresight and scientific rigor. It is not only a reinforcement of the existing legal framework for cybersecurity, data protection, and algorithmic governance, but also China's first attempt at specialized legislation addressing the new type of social relationship known as "human-machine emotional interaction." It can be seen that China's AI legislation is moving toward proactive and systematic governance, exhibiting a core thread guided by the value orientation of "responsible innovation" and the methodological approach of "systematic regulation."

I. From Regulated Objects as Tools to Systematic Guarantees for Innov\,\ldots
\end{quote}

\subsection*{gov.cn (ministry) (CN)}
\paragraph{Low-AI example (2024, \texttt{fraction\_ai}~$=$~0.00).}

\begin{quote}\small
\begin{CJK}{UTF8}{gbsn}
中国人民银行 金融监管总局 中国证监会 国家外汇局 天津市人民政府

联合发布《关于金融支持天津高质量发展的意见》

为全面贯彻落实习近平总书记关于京津冀协同发展的重要讲话和指示批示精神，强化金融对天津高质量发展的支撑作用，近日，中国人民银行、金融监管总局、中国证监会、国家外汇局、天津市人民政府联合发布《关于金融支持天津高质量发展的意见》（以下简称《意见》）。

《意见》出台着眼于提升天津金融业发展质量，营造良好金融发展环境，增强服务实体经济能力。《意见》从加强金融支持科技创新与金融科技赋能、推动金融与数字技术深度融合、完善绿色金融体系、加快普惠金融和养老金融发展、优化自贸试验区和跨境金融服务、立足地方优势促进特色金融高质量发展、加强金融监管防范化解金融风险等七个方面,提出34条重点任务和保障措施，对于做好金融“五篇大文章”，推动天津金融高质量发展具有重要意义。

下一步，中国人民银行将认真贯彻落实党的二十届三中全会精神，会同相关部门推动《意见》各项举措落实落地，为深入推进京津冀协同发展和天津加快建设社会主义现代化大都市提供有力支撑。

附件：

中国人民银行 金融监管总局 中国证监会 国家外汇局 天津市人民政府

关于金融支持天津高质量发展的意见

为全面贯彻落实习近平总书记关于京津冀协同发展的重要讲话和重要指示精神，深入推进党中央关于京津冀协同发展的战略部署，落实中央金融工作会议要求，强化金融对天津加快建设社会主义现代化大都市和经济高质量发展的支撑作用，现提出以下意见。

一、总体要求

以习近平新时代中国特色社会主义思想为指导，完整、准确、全面贯彻新发展理念，深刻把握金融工作的政治性、人民性，坚定不移走中国特色金融发展之路，深化金融供给侧结构性改革，做好金融“五篇大文章”，提高金融资源配置效率，为实体经济发展提供更高质量金融支持，更好服务京津冀协同发展。推进金融高质量发展，优化现代金融机构体系，引进、培育和打造高质量金融运营载体，提高金融资源集聚水平，做强做优金融基础设施，提升融资租赁、商业保理等特色金融发展水平，增强金融机构专业化服务能力，形成金融支持重点领域发展新模式。推进金融制度规则建设、高质量金融集聚发展、重点金融产业培育、金融产品业务优化、金融发展环境营造等领域依法合规探索实践，在金融与产业协同发展、金融与数字化融合发展上形成示范。统筹金融发展和安全，积极稳妥化解存量风险，稳妥审慎推进金融开放发展，全面加强金融监管，完善金融风险防控体系，有序有效防范化解重点领域风险，牢牢守住不发生系统性金融风险的底线。

二、加强金融支持科技创新与金融科技赋能，增强创新驱动力

（一）健全科技金融组织体系。支持金融机构完善科技型企业信贷服务机制。开展知识产权保险试点，提升保险中介在科技保险领域的服务能力。鼓励国家融资担保基金与天津中小企业信用融资担保有限公司开展再担保业务合作。证券公司可以根据区域性股权市场的业务特点和企业需求，在做好场内业务风险隔离的情况下，设立专门从事区域性股权市场相关业务的一级子公司或普惠服务部门，为“专精特新”专板企业提供投融资对接、证券承销等服务，探索形成适合场外市场特点的业务模式和管理制度。支持引进知识产权服务机构，搭建知识产权评估平台，健全知识产权融资机制。

（二）完善科技金融产品服务。大力发展知识产权质押、应用企业创新积分，加大信用贷款投放力度。有效发挥保险公司、担保机构等风险分担和增信作用，提升科技型企业首贷比，扩大科技型企业信贷覆盖面。支持商业银行、保险机构与创投股权投资机构依法合规合作，优化科技金融服务模式。支持保险机构发展科技保险业务，提供知识产权质押融资保险、专利执行和专利被侵权损失保险等新型保险产品，与中试机构合作开展相关保险业务。规范探索知识产权证券化，推动知识产权及相关实体资产组合式质押融资新模式发展。健全首台（套）重大技术装备保险、新材料首批次应用保险、软件首版次应用保险等保险补偿机制。推动高校、科研院所与区域性股权市场围绕科技成果转化开展对接合作，为创业项目发展提供早期资本市场服务。

（三）支持科技型企业利用多层次资本市场加快发展。培育天使投资人，鼓励社会资本按市场化方式设立创投基金、并购投资基金、产业投资基金等股权投资基金，建立全周期科技创新投资体系。支持天津参与实施“科技产业金融一体化”专项，引导社会资本投早、投小、投硬科技。支持符合条件的外资股权投资基金、创业投资基金落户，吸引投资周期长、资金体量大的长期资本、耐心资本落户天津。支持建设私募股权转让平台，支持二手份额转让基金（S基金）发展。研究推进优先股试点工作。支持符合条件的企业在天津设立公募基金管理公司。鼓励符合条件的科技型企业在境内外上市融资。探索完善科技型企业债券融资增信机制。全面落实区域性股权市场制度和业务创新试点，支持区域性股权市场\,\ldots
\end{CJK}
\end{quote}
\noindent\textit{English translation.}

\begin{quote}\small
People's Bank of China, National Financial Regulatory Administration, China Securities Regulatory Commission, State Administration of Foreign Exchange, and Tianjin Municipal People's Government

Jointly Issue the "Opinions on Financial Support for Tianjin's High-Quality Development"

In order to comprehensively implement the spirit of General Secretary Xi Jinping's important speeches and instructions on the coordinated development of the Beijing-Tianjin-Hebei region, and to strengthen the role of finance in supporting Tianjin's high-quality development, the People's Bank of China, the National Financial Regulatory Administration, the China Securities Regulatory Commission, the State Administration of Foreign Exchange, and the Tianjin Municipal People's Government have recently jointly issued the "Opinions on Financial Support for Tianjin's High-Quality Development" (hereinafter referred to as the "Opinions").

The Opinions were formulated with a view to improving the quality of development of Tianjin's financial industry, creating a favorable financial development environment, and enhancing the capacity to serve the real economy. The Opinions set out 34 key tasks and supporting measures in seven areas: strengthening financial support for scientific and technological innovation and financial technology empowerment; promoting deep integration of finance and digital technology; improving the green finance system; accelerating the development of inclusive finance and pension finance; optimizing free trade pilot zone and cross-border financial services; leveraging local advantages to promote the high-quality development of specialized finance; and strengthening financial supervision to prevent and defuse financial risks. The Opinions are of great significance for doing well the financial "five major articles," and for promoting the high-quality development of Tianjin's finance sector.

As the next step, the People's Bank of China will earnestly implement the spirit of the\,\ldots
\end{quote}

\subsection*{gov.cn (State Council) (CN)}
\paragraph{Low-AI example (2021, \texttt{fraction\_ai}~$=$~0.00).}

\begin{quote}\small
\begin{CJK}{UTF8}{gbsn}
国务院办公厅关于对国务院第八次大督查

发现的典型经验做法给予表扬的通报

国办发〔2021〕44号

各省、自治区、直辖市人民政府，国务院各部委、各直属机构：

为进一步推动中央经济工作会议部署和《政府工作报告》提出的目标任务落到实处，国务院部署开展了第八次大督查。从督查情况看，各有关地区在以习近平同志为核心的党中央坚强领导下，以习近平新时代中国特色社会主义思想为指导，认真贯彻党中央、国务院重大决策部署，统筹推进新冠肺炎疫情防控和经济社会发展，扎实做好“六稳”工作、全面落实“六保”任务，各项工作取得积极成效。在对16个省（自治区、直辖市）开展实地督查时发现，有关地方围绕减税降费助企发展、扩内需保就业保民生、深化“放管服”改革优化营商环境、推进创新驱动发展等方面，结合本地实际，迎难而上，勇于创新，创造和形成了一批好的经验做法。

为表扬先进，宣传典型，进一步调动和激发各方面真抓实干、改革创新的积极性、主动性和创造性，推动形成干事创业、竞相发展的良好局面，经国务院同意，对北京市坚持“一抓三保五强化”推动实现更加充分更高质量就业等48项典型经验做法予以通报表扬。希望受到表扬的地方珍惜荣誉，再接再厉，充分发挥模范表率作用，不断取得新的更大成绩。

各地区各部门要全面贯彻党的十九大和十九届二中、三中、四中、五中、六中全会精神，统筹推进“五位一体”总体布局，协调推进“四个全面”战略布局，坚持稳中求进工作总基调，立足新发展阶段，完整、准确、全面贯彻新发展理念，构建新发展格局，推动高质量发展，积极应对各种风险挑战。要学习借鉴典型经验做法，加大宣传推广力度，结合实际创造性开展工作，为完成全年经济社会发展目标任务、实现“十四五”良好开局作出积极贡献。

附件：国务院第八次大督查发现的典型经验做法（共48项）

国务院办公厅

2021年11月8日

（此件公开发布）

附件

国务院第八次大督查发现的典型经验做法

（共48项）

1．北京市坚持“一抓三保五强化”推动实现更加充分更高质量就业。

2．北京市建立完善“五新”机制高标准建设新型研发机构。

3．北京市打造“智慧税务”助推惠企利民政策落地见效。

4．天津市实施集中攻坚精准推进重点产业链高质量发展。

5．天津市优化不动产登记服务方便企业和群众办事。

6．天津市河西区多管齐下着力满足老年人养老需求。

7．河北省分类施策积极推动县域科技创新。

8．河北省石家庄市用好失业保险稳岗返还政策助企业稳就业。

9．河北省邢台市推进“医养一体、两院融合”探索农村养老新路径。

10．山西省积极打造晋字号“特”、“优”农产品品牌。

11．山西省推动社会保险数据共享提升缴费便利度。

12．山西省运城市盐湖区搭建零工市场助力务工人员灵活就业。

13．内蒙古自治区坚决遏制“两高”项目盲目发展。

14．内蒙古自治区多措并举加快推进奶业振兴发展。

15．内蒙古自治区实施“科技兴蒙”行动激发区域创新活力。

16．辽宁省以群众需求为导向努力建设人民满意的“四好农村路”。

17．辽宁省沈阳市浑南区组织“三落实”专项行动促进振兴发展提速增效。

18．辽宁省丹东市打造税务服务平台“一颗子”盘活中小微企业发展“一盘棋”。

19．吉林省长春市打造新型众创空间帮助创客实现创业梦想。

20．吉林省四平市着力推广黑土地保护利用新模式。

21．吉林省松原市前郭尔罗斯蒙古族自治县做好“加减乘除”推动查干湖实现生态保护和生态旅游双丰收。

22．安徽省合肥市系统推进综合性国家科学中心建设。

23．安徽省马鞍山市统筹推进生态环境高水平保护和产业高质量发展。

24．安徽省芜湖市实施“1\%工作法”全力推动企业降本增效。

25．江西省推行绿色金融改革促进绿色经济稳步发展。

26．江西省开展“项目大会战”扩大有效投资。

27．江西省搭建“网上中介服务超市”提供规范透明高效服务。

28．山东省加快新旧动能转换着力推动经济高质量发展。

29．山东省济南市建立企业服务中心实现惠企政策“一口办理”。

30．山东省烟台市构建一体化信用监管体系加强事中事后监管。

31．广西壮族自治区选派工业振兴特派员为企业排忧解难。

32．广西壮族自治区桂林市全力促进漓江流域生态环境持续向好。

33．广西壮族自治区柳州市精准发力推动螺蛳粉产业快速发展。

34．海南省以“机器管规划”赋能国土空间智慧治理。

35．海南省全力支持南繁科研育种基地开展种源关键核心技术攻关。

36．海南省洋浦经济开发区深化制度集成创新打造海南自由贸易港建设“样板间”。

37．重庆市开发“渝快办”平台优化市民政务服务体验。

38．重庆市开展商业价值信用贷款试点助力中小企业发展。

39．重庆市聚焦“科创+产业”打造重要创新策源地。

40．四川省突出效能导向确保财政直达资金精准有效惠企利民。

41．四川省成都市建设“科创通”服务平台激发企业创\,\ldots
\end{CJK}
\end{quote}
\noindent\textit{English translation.}

\begin{quote}\small
Circular of the General Office of the State Council on Commending Typical Experiences and Practices Discovered During the State Council's Eighth Major Inspection

Guoban Fa [2021] No. 44

People's Governments of All Provinces, Autonomous Regions, and Municipalities Directly under the Central Government; All Ministries and Commissions under the State Council; All Directly Subordinate Institutions:

In order to further drive the implementation of the arrangements made at the Central Economic Work Conference and the targets and tasks set forth in the Government Work Report, the State Council deployed and carried out its Eighth Major Inspection. Based on the inspection results, the relevant localities, under the strong leadership of the Party Central Committee with Comrade Xi Jinping as its core and guided by Xi Jinping Thought on Socialism with Chinese Characteristics for a New Era, have conscientiously implemented the major decisions and deployments of the Party Central Committee and the State Council, coordinated the prevention and control of COVID-19 with economic and social development, solidly performed the "Six Stabilizations" work and comprehensively implemented the "Six Guarantees" tasks, achieving positive results across all areas of work. During on-site inspections conducted in 16 provinces (autonomous regions and municipalities directly under the Central Government), it was found that relevant localities, in the areas of tax and fee reductions to assist enterprise development, expanding domestic demand and safeguarding employment and people's livelihoods, deepening "streamlining administration, delegating power, strengthening regulation, and improving services" (fang guan fu) reform to optimize the business environment, and advancing innovation-driven development, combined with local realities, rose to meet difficulties, dared to innovate, and created and developed a number of good experiences and practices.

In order to commend the advanced, publicize typica\,\ldots
\end{quote}

\subsection*{gov.cn (commentary) (CN)}
\paragraph{Low-AI example (2021, \texttt{fraction\_ai}~$=$~0.00).}

\begin{quote}\small
\begin{CJK}{UTF8}{gbsn}
近日召开的国务院常务会议决定，于今年7月择时启动发电行业全国碳排放权交易市场上线交易。国新办14日举行政策例行吹风会，介绍启动全国碳排放权交易市场上线交易的有关情况。

生态环境部副部长赵英民表示，目前全国碳市场相关建设任务已经基本完成，各项准备工作已经就绪。“纳入首批碳市场覆盖的企业碳排放量超过40亿吨二氧化碳，意味着中国的碳排放权交易市场一经启动，就将成为全球覆盖温室气体排放量规模最大的碳市场。”

我国的碳市场建设是从地方试点起步的。生态环境部应对气候变化司司长李高介绍，从2011年10月以来，在北京、天津、上海、重庆、湖北、广东等地开展了碳排放权交易地方试点工作，地方试点从2013年6月陆续启动了交易，经过多年发展取得了积极进展。

试点市场覆盖了电力、钢铁、水泥等20多个行业近3000家重点排放单位，到今年6月，试点省市碳市场累计配额成交量达4.8亿吨二氧化碳当量，成交额约114亿元。重点排放单位履约率保持很高水平，市场覆盖范围内碳排放总量和强度保持双降，促进了企业温室气体减排，强化了社会各界低碳发展的意识，为全国碳市场建设积累了宝贵经验。

全国碳市场选择以发电行业为突破口，有两个方面考虑：一是发电行业直接烧煤，二氧化碳排放量比较大。包括自备电厂在内的全国2000多家发电行业重点排放单位，年排放二氧化碳超过40亿吨，把发电行业作为首批启动行业，能够充分地发挥碳市场控制温室气体排放的积极作用。二是发电行业的管理制度相对健全，数据基础比较好。排放数据的准确、有效获取，是开展碳市场交易的前提。

“全国碳市场对碳达峰、碳中和的作用和意义非常重要。”赵英民介绍，主要体现在几个方面：一是推动碳市场管控的高排放行业实现产业结构和能源消费的绿色低碳化，促进高排放行业率先达峰。二是为碳减排释放价格信号，并提供经济激励机制，将资金引导至减排潜力大的行业企业，推动绿色低碳技术创新。三是通过构建全国碳市场抵消机制，促进增加林业碳汇，促进可再生能源的发展，倡导绿色低碳的生产和消费方式。四是依托全国碳市场，为行业、区域绿色低碳发展转型，实现碳达峰、碳中和提供投融资渠道。

下一步，生态环境部将推动出台《碳排放权交易管理暂行条例》，进一步完善相关的技术法规、标准、管理体系。在发电行业碳市场健康运行的基础上，逐步将市场覆盖范围扩大到更多高排放行业，根据需要丰富交易品种和交易方式，实现全国碳市场的平稳有效运行和健康持续发展。（记者 刘毅）
\end{CJK}
\end{quote}
\noindent\textit{English translation.}

\begin{quote}\small
A State Council executive meeting held recently decided to launch online trading on the national carbon emissions trading market for the power generation sector at a chosen time in July of this year. On the 14th, the State Council Information Office held a routine policy briefing to introduce the relevant circumstances surrounding the launch of online trading on the national carbon emissions trading market.

Zhao Yingmin, Vice Minister of the Ministry of Ecology and Environment, stated that the relevant construction tasks for the national carbon market have been basically completed and all preparatory work is in place. "The carbon emissions of enterprises covered in the first batch of the carbon market exceed 4 billion tonnes of CO$_2$, which means that once China's carbon emissions trading market is launched, it will immediately become the largest carbon market in the world in terms of the scale of greenhouse gas emissions covered."

China's carbon market construction started from local pilot programs. Li Gao, Director-General of the Department of Climate Change of the Ministry of Ecology and Environment, introduced that since October 2011, local pilot carbon emissions trading programs have been carried out in Beijing, Tianjin, Shanghai, Chongqing, Hubei, Guangdong, and other localities. The local pilots successively launched trading starting in June 2013, and have achieved positive progress over years of development.

The pilot markets cover nearly 3,000 key emissions units in more than 20 industries including electricity, steel, and cement. As of June of this year, the cumulative volume of quota transactions in the carbon markets of the pilot provinces and cities reached 480 million tonnes of CO$_2$ equivalent, with a transaction value of approximately 11.4 billion yuan. The compliance rate of key emissions units has remained at a very high level, and the total volume and intensity of carbon emissions within the market coverage have maintained a double decline, promoti\,\ldots
\end{quote}

\paragraph{High-AI example (2025, \texttt{fraction\_ai}~$=$~1.00).}

\begin{quote}\small
\begin{CJK}{UTF8}{gbsn}
近日，

《关于健全“高效办成一件事”重点事项常态化推进机制的意见》

正式发布，把“高效办成一件事”工作制度化、常态化，推动在更多领域更大范围加强部门协同和服务集成，将有效完善数字时代政府治理体系、提升治理水平。

一、常态化机制的战略定位：从服务创新到治理体系重构

“高效办成一件事”常态化推进机制的建立，标志着我国政务服务改革进入新阶段。这一机制绝非简单优化办事流程的技术性调整，而是数字时代政府治理体系系统性重构的关键支点。其核心价值在于将分散的行政资源整合为有机整体，通过制度化、标准化、常态化的路径，推动政府职能转变。文件首次明确将经营主体全生命周期、个人全生命周期的需求作为政务服务设计的原点，通过总体清单、年度清单、特色清单三级清单体系，实现需求精准锚定与动态响应。这种以需求侧改革倒逼供给侧优化的思路，正是治理现代化从理念到实践的重大突破。

二、制度创新的三维突破：流程再造、系统整合与标准统一

文件在制度设计上呈现出三重突破性特征。

在业务流程层面，它要求彻底打破部门藩篱，推行“一次告知、一表申请、一套材料、一窗（端）受理、一网办理”的集成服务模式，通过部门协同与互通互认实现材料精简，通过时限压缩实现效率跃升。这种革命性再造的本质，这种改变的本质在于将原本割裂的行政流程重构为以用户需求为主线的服务链条。

在系统支撑层面，文件强调全国一体化在线政务服务平台的枢纽作用，明确“省级统筹为主、地市自建为辅”的建设原则，从源头上破解系统孤岛难题。

在标准规范层面，强化“一件事”的标准统一，制定“高效办成一件事”办理模式、政务服务大厅和平台集约化建设指南等国家标准，使“无差别受理、同标准办理”成为可能。这三维突破共同构成了政府治理从碎片化走向整体性的制度基石。

三、数字时代的核心驱动：数据共享与技术赋能的深度融合

数字化手段已成为治理现代化的核心引擎，文件中将其创新性体现在三个层面。

在数据共享机制上，依托全国一体化政务大数据体系，强化政务数据共享供需对接，持续完善政务数据共享责任清单，破解“要数难、数不准”的顽疾。特别是建立垂管系统属地返还机制，大力支持基层服务创新，这种“数据回流”模式具有开创性意义。

在技术融合应用上，文件前瞻性部署人工智能等新技术与政务服务的深度融合，明确面向企业和群众、面向工作人员的不同应用场景，推动服务模式从“能办”向“智办”跃迁。

在安全与发展平衡上，文件要求提高安全防护能力、加强全流程安全管理，在确保安全的前提下，强化系统统筹建设、互联互通和政务数据共享，稳妥有序推进人工智能大模型等新技术在政务服务领域应用。

四、治理现代化的实现路径：理念转型、能力重构与生态培育

“高效办成一件事”常态化的深层价值，在于推动政府治理能力的系统性升级。这种升级首先体现为治理理念的根本转型------从管理本位转向服务本位，将群众满意度作为衡量政府绩效的核心标尺。其次推动组织架构的重构，推动跨部门协同从临时机制变为常态配置。更深层次的是公务员能力模型的重塑，要求从单一专业能力向数据素养、系统思维、协同意识等复合能力转变。

文件还特别关注数字生态培育，通过鼓励地方探索特色场景形成“一地创新、多地复用”的机制，通过保留线下服务渠道防范数字鸿沟。这种多元共治的生态构建，正是治理现代化从政府单边推进向社会协同演进的重要标志。

五、未来挑战与战略展望：在深化改革中构建新治理范式

常态化推进机制的实施仍面临打破传统藩篱、深化数字应用、释放数据价值、平衡地区行业差异等多重挑战。面向未来，应当以“高效办成一件事”为支点撬动三大变革：在服务层面发展无感申报、智能预审等新型形态，实现从“减证便民”到“无证利民”的跨越；在治理层面构建“数据驱动决策”的闭环机制，推动政策制定从经验判断向精准施策转变；在发展层面探索“数字政府---数字经济”协同模式，通过开放场景培育创新生态。唯有将常态化机制内化为政府运行的底层逻辑，才能真正建成“泛在可及、智慧便捷、公平普惠”的数字政府，为中国式现代化提供坚实的治理支撑。

（清华大学公共管理学院教授 孟庆国）
\end{CJK}
\end{quote}
\noindent\textit{English translation.}

\begin{quote}\small
Recently, the "Opinions on Establishing a Normalized Promotion Mechanism for Key Matters in 'Efficiently Completing Any Given Task'" was officially issued, institutionalizing and normalizing the work of "efficiently completing any given task," and driving the strengthening of inter-departmental coordination and service integration across more domains and broader areas. This will effectively improve the government governance system in the digital era and raise the level of governance.

I. Strategic Positioning of the Normalized Mechanism: From Service Innovation to Reconstruction of the Governance System

The establishment of the normalized promotion mechanism for "efficiently completing any given task" marks the entry of China's government services reform into a new stage. This mechanism is by no means a mere technical adjustment to optimize administrative procedures; it is a key fulcrum for the systemic reconstruction of the government governance system in the digital era. Its core value lies in integrating dispersed administrative resources into an organic whole, and advancing the transformation of government functions through institutionalized, standardized, and normalized pathways. For the first time, the document explicitly places the needs of business entities throughout their full life cycle and individuals throughout their full life cycle as the starting point for government service design. Through a three-tier list system comprising an overall list, an annual list, and a featured list, it achieves precise anchoring of demand and dynamic responsiveness. This approach of using demand-side reform to compel supply-side optimization is a major breakthrough in the modernization of governance from concept to practice.

II. Three-Dimensional Breakthroughs in Institutional Innovation: Process Reengineering, System Integration, and Standard Unification

The document exhibits three breakthrough characteristics in institutional design.

At the level of business process\,\ldots
\end{quote}

\paragraph{High-AI example (2025, \texttt{fraction\_ai}~$=$~0.54).}

\begin{quote}\small
\begin{CJK}{UTF8}{gbsn}
一、制定背景和目的

自2010年起，我委先后制定《电子病历基本规范（试行）》《电子病历应用管理规范（试行）》《电子病历系统功能规范（试行）》《医疗机构临床决策支持系统应用管理规范（试行）》等文件，明确电子病历建立、管理要求及系统功能标准，强化技术及质量要求，维护医患双方的合法权益，持续推进以电子病历为核心的医疗机构信息化建设。为进一步加强医疗机构和医务人员管理，规范患者医疗信息使用，国家卫生健康委会同国家中医药局、国家疾控局制定

《关于进一步加强医疗机构电子病历信息使用管理的通知》

（以下简称《通知》），旨在通过压实医疗机构主体责任，强化监管措施，进一步保障患者医疗信息和医疗质量安全。

二、《通知》内容

（一）加强医疗机构内部管理。医疗机构需明确电子病历范围，压实主体责任，依法依规严格保护患者隐私，将电子病历信息规范使用管理情况纳入绩效评价。健全管理制度，建立电子病历使用长效监管机制和应急处置制度。落实分级管理要求，遵循最小可用原则，明确临床诊疗、教学、管理等相关人员分级访问权限和时限。

（二）规范电子病历信息使用。医疗机构需规范相关人员使用权限和行为，不得违规收集、传输或泄露患者信息。加强短期人员培训与管理，确保权限与职责匹配，并与外部服务商签订保密协议。保障全流程可追溯，采用数字水印等技术，确保使用过程留痕。确保数据安全，建立电子病历信息安全防护体系，防范潜在安全风险。

（三）强化卫生健康行政部门监管。地方各级卫生健康行政部门(含中医药、疾控部门，下同)要加强对医疗机构指导和监管，组织推进落实，定期监测评估。各省级卫生健康行政部门要将医疗机构规范使用电子病历信息情况作为医院评审、医院巡查、智慧医院建设等相关工作重要评估依据。

三、抓好文件落实

各地要按照《通知》要求结合实际抓好落实，做好教育培训和政策解读，定期对医疗机构电子病历信息的使用管理情况进行监测和评估，规范医疗机构电子病历信息使用，保障患者医疗信息和医疗质量安全。
\end{CJK}
\end{quote}
\noindent\textit{English translation.}

\begin{quote}\small
I. Background and Purpose of Formulation

Beginning in 2010, our Commission successively formulated documents including the Basic Standards for Electronic Medical Records (Trial), the Standards for the Administration of Electronic Medical Record Applications (Trial), the Functional Standards for Electronic Medical Record Systems (Trial), and the Standards for the Administration of Clinical Decision Support System Applications in Medical Institutions (Trial), which clarified the requirements for the establishment and management of electronic medical records as well as system functional standards, strengthened technical and quality requirements, safeguarded the lawful rights and interests of both doctors and patients, and continuously advanced the informatization of medical institutions centered on electronic medical records. In order to further strengthen the management of medical institutions and medical personnel, and to standardize the use of patients' medical information, the National Health Commission, together with the National Administration of Traditional Chinese Medicine and the National Disease Control and Prevention Administration, has formulated the Notice on Further Strengthening the Administration of the Use of Electronic Medical Record Information in Medical Institutions (hereinafter referred to as the "Notice"), which aims to further safeguard the security of patients' medical information and medical quality by consolidating the primary responsibilities of medical institutions and strengthening supervisory measures.

II. Contents of the Notice

(1) Strengthening internal management of medical institutions. Medical institutions are required to clarify the scope of electronic medical records, consolidate primary responsibilities, strictly protect patient privacy in accordance with laws and regulations, and incorporate the standardized use and management of electronic medical record information into performance evaluations. They shall improve management\,\ldots
\end{quote}

\subsection*{MOST (CN)}
\paragraph{Low-AI example (2025, \texttt{fraction\_ai}~$=$~0.00).}

\begin{quote}\small
\begin{CJK}{UTF8}{gbsn}
11月17日上午，央地共建成渝地区区域科技创新中心工作推进会在重庆召开。重庆市委书记袁家军出席会议并讲话。科技部党组书记、部长阴和俊，四川省委副书记、省长施小琳讲话。重庆市委副书记、市长胡衡华主持。

会上，重庆市、四川省负责同志分别通报了推进成渝地区区域科技创新中心建设进展及下一步工作安排。科技部负责同志介绍了新时期推进成渝地区区域科技创新中心建设的顶层设计文件及落实情况。中央和国家有关部门介绍了支持成渝地区区域科技创新中心建设的工作情况及2026年工作考虑。

袁家军在讲话中强调，建设成渝地区区域科技创新中心，是以习近平同志为核心的党中央作出的重大战略部署，是川渝两省市共同肩负的重大责任。近年来，重庆深入学习贯彻习近平总书记关于科技创新的重要论述，以成渝地区双城经济圈建设为总牵引，深化实施科技创新和人才强市首位战略，川渝联动构建完善区域协同创新体系，推动成渝地区区域科技创新中心建设取得新进展新成效。新征程上，重庆将全面学习贯彻党的二十届四中全会精神，牢固树立川渝“一盘棋”思想和一体化发展理念，坚持以科技创新引领构建现代化产业体系、激发新动能，突出“人工智能+”和深化科技体制机制改革“双轮驱动”，推动共建成渝地区区域科技创新中心取得更大标志性成果。要全力壮大战略科技力量，打造紧密型创新共同体，共建高水平实验室体系，引育构建产业技术创新平台矩阵。要加快建设人工智能应用高地，强化人工智能赋能科创模式变革、资源配置、生态优化，丰富拓展示范性标志性应用场景，持续提升“数字成渝”共建水平。要提质打造具有国际竞争力的产业创新策源地，聚焦智能网联新能源汽车、新一代电子信息、先进材料等重点产业补短板强优势，不断提升现代化产业体系发展能级。要努力营造国际一流创新生态，高水平培育创新人才队伍，提高“双一流”建设质效，强化科技金融支撑，构建创新成果全链条孵化体系，持续深化科技开放合作。要健全成渝协同机制，抓好要素配套保障，推动重点任务和合作事项闭环落实，进一步凝聚共建成渝地区区域科技创新中心强大合力。

阴和俊指出，建设成渝地区区域科技创新中心是党中央赋予川渝两地的共同战略使命。习近平总书记亲自谋划、亲自部署、亲自推动成渝地区区域科技创新中心建设，对成渝地区科技创新作出系列重要指示，为成渝地区区域科技创新中心建设工作提供了根本遵循。在央地共同努力下，成渝地区区域科技创新中心建设取得积极成效，战略科技力量持续增强，关键核心技术攻关实现新突破，川渝协同创新迈出新步伐，为科技强国建设提供了坚实支撑。成渝地区科教资源丰富、产业体系完善、创新潜力大，要在发展新质生产力、开展基础前沿攻关、营造创新生态、扩大开放合作等方面唱好“双城记”，在西部大开发新格局中一道“打头阵”，在科技强国建设中一起“挑大梁”，进一步提升全国影响力、全球竞争力。科技部将认真学习贯彻习近平总书记重要指示精神，进一步强化央地协同、部门联动，充分发挥重庆市、四川省主体作用，以更大力度、更实举措支持成渝地区区域科技创新中心强化协同联动，凝聚形成强大工作合力；推动科技创新和产业创新深度融合，聚力突破一批关键核心技术；壮大战略科技力量，提升原始创新策源能力；统筹推进教育科技人才一体改革，构建支持全面创新体制机制；营造开放创新环境，打造具有国际竞争力的创新生态；建立健全工作机制，强化统筹协调，支持成渝地区加快建成具有全国影响力的科技创新中心，辐射带动周边地区高质量发展，为科技强国建设作出新的更大贡献。

施小琳指出，建设成渝地区区域科技创新中心，是以习近平同志为核心的党中央作出的重大战略部署，是服务国家高水平科技自立自强的重大任务、打造科技强国建设重要战略支点的关键支撑，也是川渝推动高质量发展的重要牵引。我们将牢记习近平总书记赋予成渝的“一极一源、两中心两地”和赋予四川的“两高地、两基地、一屏障”使命任务，深入贯彻党的二十届四中全会精神，在科技部等国家部委和重庆市支持帮助下，站位全局、真抓实干，高效运行跨区域协同创新机制，推动产业教育科技人才一体发展，培育形成一批万亿级创新型产业集群，加快发展壮大新质生产力，提升科技创新治理体系和治理能力，着力建设具有全国影响力的科技创新中心。坚持科技创新打头阵，聚合资源、向新图强，强化科技创新策源功能，构建多层次实验室体系，实施西部数据基础设施等新基建工程，抢抓国家采取超常规措施全链条推动重点领域关键核心技术攻关取得决定性突破重大机遇，争取承担更多重大科技任务，大力推动科技成果转化应用，促进“四链”深度融合，加快建设西部地区创新高地。树牢一盘棋思想和一体化发展理念，完善机制、协同联动，深化成渝绵“创新金三角”和成渝中线科创走廊建设，设立实施合作专项，推动“一带一路”科技交流大会成果落地转化，加强知识产权保护运用，协同完善科技服务体系，携手创造一流创新生态、打造川渝创新共\,\ldots
\end{CJK}
\end{quote}
\noindent\textit{English translation.}

\begin{quote}\small
On the morning of November 17, the Work Advancement Meeting for the Central-Local Co-Construction of the Chengdu-Chongqing Regional Science and Technology Innovation Center was held in Chongqing. Yuan Jiajun, Secretary of the Chongqing Municipal Party Committee, attended and delivered remarks. Yin Hejun, Party Secretary and Minister of the Ministry of Science and Technology, and Shi Xiaolin, Deputy Secretary of the Sichuan Provincial Party Committee and Governor, delivered remarks. Hu Henghua, Deputy Secretary of the Chongqing Municipal Party Committee and Mayor, presided over the meeting.

At the meeting, responsible officials from Chongqing and Sichuan respectively briefed attendees on the progress of the Chengdu-Chongqing Regional Science and Technology Innovation Center construction and the arrangements for the next steps. Responsible officials from the Ministry of Science and Technology introduced the top-level design documents for advancing the construction of the Chengdu-Chongqing Regional Science and Technology Innovation Center in the new era and the status of their implementation. Relevant central and national departments introduced their work in supporting the construction of the Chengdu-Chongqing Regional Science and Technology Innovation Center and their work considerations for 2026.

In his remarks, Yuan Jiajun emphasized that building the Chengdu-Chongqing Regional Science and Technology Innovation Center is a major strategic deployment by the Party Central Committee with Comrade Xi Jinping as the core, and is a major responsibility jointly shouldered by the two provinces and municipalities of Sichuan and Chongqing. In recent years, Chongqing has studied and implemented in depth General Secretary Xi Jinping's important expositions on science and technology innovation, using the construction of the Chengdu-Chongqing Twin-City Economic Circle as the overall driver, deepening the implementation of the primary strategy of science and technology innovation\,\ldots
\end{quote}

\paragraph{High-AI example (2026, \texttt{fraction\_ai}~$=$~0.64).}

\begin{quote}\small
\begin{CJK}{UTF8}{gbsn}
在科技日报创刊40周年之际，中共中央总书记、国家主席、中央军委主席习近平向科技日报发来贺信，向报社全体同志表示祝贺。1月1日上午，庆祝科技日报创刊40周年大会在北京举行。科技部党组书记、部长阴和俊宣读习近平总书记致科技日报创刊40周年贺信并讲话。

阴和俊在讲话中指出，习近平总书记就科技日报创刊40周年专门发来贺信，充分肯定科技日报围绕党和国家中心工作发挥的积极作用，对在新起点上做好科技宣传工作提出明确要求。这体现了对科技创新、科技宣传工作的高度重视，以及对科技日报社的关心厚爱和殷切期望，为进一步办好科技日报、加强科技宣传、做好科技工作指明了前进方向，提供了根本遵循。

阴和俊表示，40年来，科技日报始终坚持党的领导，切实履行科技宣传职责使命，为我国科技事业和经济社会发展营造了良好舆论氛围。他强调，科技日报要坚持正确政治方向，深入宣传阐释习近平新时代中国特色社会主义思想；精心讲好中国科技创新故事，切实增强全社会的创新自信；大力弘扬科学家精神，持续培育创新文化；积极反映科技工作者心声，为科技事业发展凝心聚力；创新载体方式，不断提升传播效能和影响力。

科技部党组成员、科技日报社社长吴兢在致辞中回顾了科技日报40年走过的不凡历程。她表示，要深入学习贯彻习近平总书记致科技日报创刊40周年贺信精神，把守正作为立身之本，把唯实作为成事之基，把务新作为发展之要，更好发挥党和国家科技宣传主阵地作用，努力开创科技宣传事业新局面，为科技强国建设贡献新的更大力量。

大会由科技部党组成员、副部长林新主持，中央纪委国家监委驻科学技术部纪检监察组组长、科技部党组成员高波，科技部党组成员、秘书长潘晓东，科技部副秘书长、六司司长苗鸿，中央和国家机关有关部门和单位、中央新闻单位、地方宣传部门负责人等参加大会。
\end{CJK}
\end{quote}
\noindent\textit{English translation.}

\begin{quote}\small
On the occasion of the 40th anniversary of the founding of Science and Technology Daily, Xi Jinping, General Secretary of the CPC Central Committee, President of the State, and Chairman of the Central Military Commission, sent a congratulatory letter to Science and Technology Daily, extending congratulations to all colleagues at the newspaper. On the morning of January 1st, a conference celebrating the 40th anniversary of the founding of Science and Technology Daily was held in Beijing. Yin Hejun, Secretary of the Party Leadership Group and Minister of the Ministry of Science and Technology, read aloud General Secretary Xi Jinping's congratulatory letter to Science and Technology Daily on its 40th founding anniversary and delivered a speech.

In his speech, Yin Hejun noted that General Secretary Xi Jinping had sent a special congratulatory letter on the occasion of the 40th anniversary of the founding of Science and Technology Daily, fully affirming the positive role Science and Technology Daily has played in serving the central work of the Party and the state, and setting out clear requirements for doing science and technology publicity work well from a new starting point. This reflects the high degree of importance attached to science and technology innovation and science and technology publicity work, as well as the care, affection, and earnest expectations held for Science and Technology Daily, and it charts the direction forward and provides the fundamental guide for further improving Science and Technology Daily, strengthening science and technology publicity, and doing science and technology work well.

Yin Hejun stated that over the past 40 years, Science and Technology Daily has consistently upheld the Party's leadership, faithfully fulfilled its mission and responsibilities in science and technology publicity, and created a favorable public opinion environment for China's science and technology enterprise and economic and social development. He emphasized\,\ldots
\end{quote}

\paragraph{High-AI example (2026, \texttt{fraction\_ai}~$=$~1.00).}

\begin{quote}\small
\begin{CJK}{UTF8}{gbsn}
2026年4月7日，中国---黑山科技合作委员会第五届例会以视频形式在北京和波德戈里察举行。中国科技部国际合作司副司长王晓与黑山教育、科学和创新部国际合作、欧洲一体化和欧盟基金司司长马尔科·武卡希诺维奇共同主持会议。

本届例会通报了两国科学、技术和创新政策最新进展，总结了第四届例会以来中黑科技合作总体情况，就未来中黑科技创新合作的优先方向、重点任务等达成共识。

王晓表示，在两国元首的战略引领下，近年来中黑科技合作稳步发展，双方科研机构和人员交流日益频繁，项目合作不断取得新进展新成果。下一步，中方愿与黑方一道，继续深化在人员交流、联合研发、平台建设等领域的务实合作，为中黑关系发展注入新动力。

马尔科·武卡希诺维奇高度评价双方合作，表示黑方愿与中方巩固科技合作委员会机制，在科研人员往来、联合项目实施等领域继续扩大互利合作，推动双边科技合作提质升级。

科技部国际合作司、中国科学技术交流中心、中国农业大学、中国---中东欧国家创新合作研究中心等中方单位的代表参加会议。
\end{CJK}
\end{quote}
\noindent\textit{English translation.}

\begin{quote}\small
On April 7, 2026, the Fifth Session of the China--Montenegro Joint Committee on Science and Technology Cooperation was held via video link between Beijing and Podgorica. The meeting was co-chaired by Wang Xiao, Deputy Director-General of the Department of International Cooperation of China's Ministry of Science and Technology, and Marko Vuka?inovi?, Director-General of the Department of International Cooperation, European Integration and EU Funds of Montenegro's Ministry of Education, Science and Innovation.

The session briefed both sides on the latest developments in science, technology, and innovation policy in the two countries, reviewed the overall status of China--Montenegro science and technology cooperation since the Fourth Session, and reached consensus on priority directions and key tasks for future China--Montenegro science and technology innovation cooperation.

Wang Xiao stated that, under the strategic guidance of the two countries' heads of state, China--Montenegro science and technology cooperation has advanced steadily in recent years, with exchanges between research institutions and personnel becoming increasingly frequent and project cooperation continuously yielding new progress and results. Going forward, the Chinese side stands ready to work together with the Montenegrin side to continue deepening practical cooperation in areas such as personnel exchanges, joint research and development, and platform building, so as to inject new momentum into the development of China--Montenegro relations.

Marko Vuka?inovi? spoke highly of the cooperation between the two sides, stating that the Montenegrin side is willing to work with the Chinese side to consolidate the Joint Committee on Science and Technology Cooperation mechanism, continue to expand mutually beneficial cooperation in areas such as exchanges between scientific researchers and implementation of joint projects, and advance the quality improvement and upgrading of bilateral science and technology c\,\ldots
\end{quote}

\subsection*{PLA Daily (CN)}
\paragraph{Low-AI example (2025, \texttt{fraction\_ai}~$=$~0.00).}

\begin{quote}\small
\begin{CJK}{UTF8}{gbsn}
【光明论坛】

作者：完颜平（纪检监察工作者）

习近平总书记在二十届中央纪委四次全会上发表重要讲话强调，新时代新征程，必须坚持用改革精神和严的标准管党治党，努力取得更大成效，确保党的二十大和二十届三中全会部署落地落实，确保党始终成为中国特色社会主义事业的坚强领导核心，推动中国式现代化行稳致远。

推进全面从严治党取得更大成效，必须聚焦“两个维护”强化政治监督，严明政治纪律和政治规矩，以有力监督保障改革顺利推进。政治监督是实现党的政治路线的重要保障，是督促全党坚持党中央集中统一领导的有力举措。进一步全面深化改革，必须依靠有力监督来清障护航。改革越是向纵深推进，对监督保障的要求就越高。纪检监察机关是推进党的自我革命的重要力量，在推进中国式现代化进程中肩负重要使命，必须切实提高政治站位，始终心怀“国之大者”，立足职能职责，积极担当作为。围绕党的二十届三中全会部署的全面深化改革各项任务，强化政治监督，纠正政治偏差。推进政治监督具体化、精准化、常态化，聚焦各地区各部门各单位完整准确全面贯彻新发展理念、加快构建新发展格局、着力推动高质量发展等重大战略部署落实，确保各项决策部署落实到位。把全会重大改革落实情况纳入监督检查和巡视巡察内容，严明政治纪律和政治规矩，压紧压实改革责任，推动各个地区、各条战线各项改革任务落地见效。

推进全面从严治党取得更大成效，必须巩固深化党纪学习教育成果，综合发挥党的纪律教育约束、保障激励作用。必须深化运用党纪学习教育成功经验，推动党的纪律建设走深走实。深入学习领会习近平总书记关于全面加强党的纪律建设的重要论述，原原本本学习新修订的《中国共产党纪律处分条例》，以学纪知纪明纪促进遵纪守纪执纪，通过学习党的纪律，进一步锤炼党性、提高政治觉悟和道德修养，把纪律要求转化为内在追求，养成遵规守纪的自觉，在遵守和维护党的纪律上取得新进步。建立经常性和集中性相结合的纪律教育机制，坚持融入日常、抓在经常，统筹抓好纪律教育、纪律完善、纪律执行，把纪律教育融入日常教育管理监督、贯穿干部成长全周期。与时俱进完善纪律规章，及时把全面从严治党理论和实践成果上升为制度规范、转化为纪律要求。动真碰硬强化纪律执行，坚持党性党风党纪一起抓，深化运用监督执纪“四种形态”，把严的基调、严的措施、严的氛围长期坚持下去。综合发挥党的纪律教育约束、保障激励作用，充分调动干部干事创业积极性，做到廉而有为。把纪律挺在前面，不是要把人管死，而是要管好用活，促进从严执纪和鼓励担当作为相统一，营造积极健康、干事创业的良好环境。通过严纪行、立规矩、正风气、强免疫，使“能人”能“干事”，又不“出事”。把党纪学习教育成果持续转化为推动高质量发展的强大动力，让改革发展的动力活力竞相迸发。

推进全面从严治党取得更大成效，必须健全不正之风和腐败问题同查同治机制，着力推动正风反腐一体深化。风腐同查同治是我们党在新时代新征程对正风肃纪反腐实践规律的深刻认识和科学把握，是针对风腐一体同生的本质特征作出的必然选择。大量案例表明，不正之风与腐败问题同根同源、互为表里，“四风”是腐败滋生的温床，腐败又反过来催生助长“四风”。党的十八大以来，全面从严治党从改进作风破题，锲而不舍落实中央八项规定精神，经过持续整治，发生在群众眼皮底下的公款享乐奢靡问题明显减少，但一些问题仍然存在。深入推进风腐同查同治，“由腐纠风”“由风防腐”“风腐同查同治”，拓展了正风肃纪反腐的深度、广度。既要在“查”上动真格，又要在“治”上见真章，精准查找本地区本领域哪些不正之风与腐败关系最为紧密，哪些作风问题容易演变为腐败，有针对性地进行整治。深化以案促改、以案促治，大力查处腐败背后的不正之风，完善通报曝光机制，加大对风腐一体、风腐交织问题通报力度，增强震慑效应。

推进全面从严治党取得更大成效，必须持续推动全面从严治党向基层延伸，深化整治群众身边不正之风和腐败问题，推动改革发展成果更好更公平惠及广大人民群众。老百姓看全面从严治党的成效，不仅看打了多少“老虎”，也看拍了多少“苍蝇”，更多的是看身边的党员干部“正派不正派、干事不干事、干净不干净”。“蝇贪蚁腐”，官可能不大，但影响党员干部在人民群众中的形象；钱可能不多，但侵占的是人民群众的利益。从这个意义说，“微腐”并不“微”，危害同样“大”。人民群众反对什么、痛恨什么，就要坚决防范和纠正什么。二十届中央纪委二次、三次全会都对坚决惩治群众身边腐败问题作出具体部署，瞄准就业创业、教育医疗、养老社保、生态环保、安全生产、食品药品安全、执法司法等群众反映强烈的突出问题开展集中整治，解决了一批群众身边的实际问题，人民群众切实感受到公平正义就在身边。纵深推动反腐败斗争向基层延伸、向群众身边延伸，持续深化整治群众身边不正之风和腐败问题，以正风肃纪反腐新成效取信于民，不断厚植党长期执政的政治根基，\,\ldots
\end{CJK}
\end{quote}
\noindent\textit{English translation.}

\begin{quote}\small
[Guangming Forum]

Author: Wanyan Ping (discipline inspection and supervision worker)

General Secretary Xi Jinping delivered an important speech at the Fourth Plenary Session of the 20th Central Commission for Discipline Inspection, emphasizing that on the new journey in the new era, we must adhere to managing and governing the Party with a spirit of reform and strict standards, strive for greater results, ensure that the deployments of the 20th National Congress of the Communist Party of China and the Third Plenary Session of the 20th Central Committee are carried out and implemented, ensure that the Party always remains the strong leadership core of the cause of socialism with Chinese characteristics, and promote the steady and far-reaching advance of Chinese-style modernization.

To achieve greater results in advancing comprehensive and strict Party governance, we must focus on the "Two Upholds," strengthen political oversight, strictly enforce political discipline and political rules, and use robust oversight to guarantee the smooth advancement of reform. Political oversight is an important guarantee for realizing the Party's political line and a powerful measure for urging the entire Party to uphold the centralized and unified leadership of the Party Central Committee. Further comprehensively deepening reform must rely on strong oversight to clear obstacles and provide safeguards. The deeper reform advances, the higher the requirements for oversight and guarantee. Discipline inspection and supervision organs are an important force in advancing the Party's self-revolution; they bear an important mission in the process of advancing Chinese-style modernization and must genuinely raise their political standing, always keep the "big picture of the nation" in mind, and actively take responsibility in accordance with their functions and duties. Centered on the various comprehensive deepening reform tasks deployed at the Third Plenary Session of the 20th Central Commit\,\ldots
\end{quote}

\paragraph{High-AI example (2026, \texttt{fraction\_ai}~$=$~1.00).}

\begin{quote}\small
\begin{CJK}{UTF8}{gbsn}
日前，中国军号刊发的一篇带兵骨干心得，引发了不少关注与讨论。文中，这位带兵人反思自己过去总盯着战士们的朋友圈“找思想问题、查苗头隐患”，动不动就长篇大论地教育，结果不仅没能与官兵打成一片，反而把自己隔离在“圈外”。这篇心得之所以能引发共鸣，正是因为戳中了基层教育管理中的痛点。“朋友圈”是青年官兵记录生活、表达情绪的园地，但并不等同于一个人的现实全貌。一些带兵人，不擅长真正有效的方式方法，简单把“朋友圈”当成了“监视器”，一见风吹草动便说教不断，自然被官兵敬而远之。政治工作是做人的工作，做好人的工作，首先要尊重人、理解人。带兵人不妨多一些换位思考，少一些死板说教，尊重官兵的情绪表达，接纳年轻人的“网言网语”。只有真正与官兵交心交友，赢得他们的信任与尊重，才能让思想政治工作落地见效。
\end{CJK}
\end{quote}
\noindent\textit{English translation.}

\begin{quote}\small
Recently, an article sharing the reflections of a troop-leading cadre, published by China Military Network (Zhongguo Junhao), sparked considerable attention and discussion. In the article, the troop leader reflects on how he used to fixate on soldiers' Moments feeds to "look for ideological problems and check for early warning signs," readily launching into lengthy lectures at the slightest provocation --- with the result that he not only failed to integrate with the officers and soldiers, but instead found himself isolated "outside the circle." The reason this piece struck such a chord is precisely that it hit on a pain point in grassroots education and management. "Moments" is a space where young officers and soldiers record their lives and express their emotions, but it is by no means equivalent to a complete picture of a person's reality. Some troop leaders, lacking genuinely effective methods and approaches, have simply turned "Moments" into a "surveillance camera" --- the moment they detect the slightest stir, they launch into nonstop moralizing, and naturally find themselves held at arm's length by the troops. Political work is work done with people; to do it well, one must first respect people and understand people. Troop leaders would do well to practice more empathy and less rigid lecturing, to respect the emotional expression of officers and soldiers, and to accept young people's "internet speak." Only by truly forming genuine bonds of friendship and trust with officers and soldiers --- earning their trust and respect --- can ideological and political work take root and bear fruit.
\end{quote}

\paragraph{High-AI example (2026, \texttt{fraction\_ai}~$=$~1.00).}

\begin{quote}\small
\begin{CJK}{UTF8}{gbsn}
美国和以色列对伊朗发动联合军事打击，致伊最高领袖哈梅内伊在空袭中遇害。这是继今年1月突袭委内瑞拉、强行带走委总统马杜罗后，美国在不到两个月内，再次对主权国家领导人采取极端军事行动。从拉美到中东，从强行控制到公然击杀，国际秩序的边界被不断突破，人类文明的底线被公然践踏，其破坏性值得国际社会高度警惕。

将主权国家领导人作为军事打击目标，违反《联合国宪章》所确立的基本原则，是对战后数十年努力构建起来的国际规则体系的粗暴践踏。这已不是一般意义上的地缘政治博弈，而是涉及国际法底线存续的基本问题。

这一极端做法非常危险。一个军事强国绕开联合国安理会，单方面对他国领导人进行“审判”或“击杀”，这种挑衅国际法红线的逻辑一旦被默许，与其存在利益分歧的国家都可能面临类似的安全威胁，国际关系将被“武力至上”的丛林法则所取代。当武力成为应对分歧的优先选项，外交谈判和多边协商的空间必然受到严重挤压，国际争端的和平解决机制将被严重削弱。

滥用武力犹如饮鸩止渴，不仅无法解决复杂的国际分歧，反而可能引发难以预料的系统性灾难。强推政权更迭最终带来的往往是仇恨的深渊。历史殷鉴不远，正如当年在伊拉克、利比亚留下的满目疮痍与无休止的血腥内乱，今日降临在伊朗的炮火，只会将整个中东地区推向更加剧烈的动荡与撕裂。当前，海湾地区面临新一轮紧张局势升级和外溢风险，全球能源供应和地区安全格局都可能受到冲击。

世界绝不能退回到弱肉强食的“黑暗丛林”，维护以联合国为核心的国际体系和《联合国宪章》宗旨原则，关乎所有国家的共同利益。面对单边武力行动不断突破底线的趋势，国际社会应发出明确清晰声音，呼吁各方立即停止军事行动，防止局势轮番升级，通过对话谈判解决矛盾分歧，推动中东地区早日恢复和平稳定。这既是对当下危机的回应，也是对持久和平的追求。

（新华社北京3月5日电）
\end{CJK}
\end{quote}
\noindent\textit{English translation.}

\begin{quote}\small
The United States and Israel launched a joint military strike against Iran, killing Supreme Leader Khamenei in the air attack. This marks the second time in less than two months that the United States has taken extreme military action against the leader of a sovereign state, following the surprise raid on Venezuela in January of this year that forcibly removed President Maduro. From Latin America to the Middle East, from forcible seizure to brazen assassination, the boundaries of the international order are being continuously broken through and the baseline of human civilization is being openly trampled upon --- the destructive consequences of this warrant the highest vigilance from the international community.

Targeting the leader of a sovereign state for military strikes violates the fundamental principles established by the United Nations Charter and constitutes a brutal trampling of the system of international rules painstakingly constructed over the postwar decades. This is no longer a geopolitical contest in the ordinary sense; it is a fundamental question concerning the survival of the baseline of international law.

This extreme course of action is extremely dangerous. When a militarily powerful state bypasses the UN Security Council and unilaterally "tries" or "kills" the leader of another country, once the logic of provoking the red lines of international law is tacitly permitted, all countries that have differences of interest with it may face similar security threats, and international relations will be replaced by the law of the jungle in which "might makes right." When force becomes the preferred option for addressing disputes, the space for diplomatic negotiation and multilateral consultation will inevitably be severely compressed, and the mechanisms for the peaceful resolution of international disputes will be gravely weakened.

The abuse of force is like drinking poison to quench one's thirst: not only does it fail to resolve complex international di\,\ldots
\end{quote}

\subsection*{DARPA news (US)}
\paragraph{Low-AI example (2024, \texttt{fraction\_ai}~$=$~0.00).}

\begin{quote}\small
The Department of Defense (DOD) needs the flexibility to manufacture critical structures at the time and point of need using locally available materials. Yet current manufacturing approaches for forward production operate under the assumption that pristine raw materials will be readily available; furthermore, any change in material input would require costly redesign. This means that structure designs are reliant on fixed inputs to produce fixed outputs, hindering the flexibility required for forward production where raw materials and resources are limited.

DARPA's new Rubble to Rockets (R

2

) program aims to overcome current limitations to manufacturing in supply chain-denied environments by developing production and design approaches that can accommodate widely variable input materials. Performers will focus on creating an inexpensive, flexible, and robust platform for the production and characterization of raw material for use in structural fabrication. They will then seek to apply that platform to adaptively update a sounding rocket's structural design.

``Existing manufacturing approaches require stability. Building a framework that can enable the manufacture of structures out of anything, anywhere, and at many sizes would break the status quo for manufacturing in resource-contested environments,'' said Hunter Martin, DARPA's program manager for the R

2

program. ``We're focusing on sounding rockets for proof of concept because they represent a single-use structure with multiple components and complex structural requirements, but anticipate broad applicability to a wide range of manufacturing use cases -- from spare parts and infrastructure repair to system production.''

R

2

will also look to leverage material informatics and innovative processing and manufacturing techniques to dramatically drive down the timeframes and scale needed for production.

DARPA hypothesizes that the analytical framework R

2

aims to develop would allow for rapid upgrades to incorporate the g\,\ldots
\end{quote}

\paragraph{High-AI example (2025, \texttt{fraction\_ai}~$=$~1.00).}

\begin{quote}\small
Quantum sensors, which have demonstrated unmatched precision in measuring magnetic fields, gravity, and motion, hold immense promise for advancing defense capabilities. However, these sensors face significant challenges when deployed in real-world settings -- especially when placed on moving platforms. Vibrations, electromagnetic interference, and other environmental disturbances degrade their performance, limiting their operational utility. Current solutions, such as isolating the sensors or using bulky shielding, are impractical for widespread deployment.

To address these barriers, DARPA is launching the

Robust Quantum Sensors (RoQS) program

. RoQS will seek to develop quantum sensors that are inherently resistant to environmental disruptions, ensuring they can operate reliably outside of a laboratory without sacrificing their sensitivity. The ultimate goal of the program is to integrate these sensors into Department of Defense platforms.

RoQS emphasizes early collaboration between sensor developers and platform makers within the defense industrial base to streamline the transition from research to operational deployment. By fostering these partnerships early on, DARPA aims to ensure that the developed technologies meet real-world requirements and can be seamlessly integrated into existing systems -- significantly reducing the time and resources needed to harden and deploy quantum sensors.

``Quantum sensors have the potential to redefine how we gather critical information in defense scenarios,'' said

Dr. Jonathan Hoffman, Program Manager in DARPA's Microsystems Technology Office.

``However, their fragility has been a major barrier to deployment. With RoQS, we're taking a bold step toward creating sensors that are not only extremely precise but also resilient in the face of real-world challenges.''

The RoQS program is designed to tackle the problem of sensor fragility at its core. Instead of relying on bulky shielding or isolation techniques that are impractical for wide\,\ldots
\end{quote}

\paragraph{High-AI example (2026, \texttt{fraction\_ai}~$=$~0.66).}

\begin{quote}\small
In an era of unprecedented biological data generation, the ability to rapidly determine the origin of a biological event---whether natural, accidental, or intentional---is a critical component of national security and public health. To meet the challenge of finding the "needle in a haystack" within this data deluge, the Defense Advanced Research Projects Agency (DARPA) has launched the Bio-Attribution Challenge.

This virtual competition calls on innovators to develop a new generation of tools capable of analyzing petabyte-scale datasets in near real-time, far exceeding the capacity of current systems. The goal is to revolutionize how we identify and trace the source of biological sequences, ensuring a faster, more effective response to potential threats.

"The ability to rapidly and accurately identify the source of a biological sequence, whether natural or engineered, is a critical national security capability," said

Abhishek Singharoy, Ph.D.

, program manager for the Bio Attribution Challenge. ``We're calling on creative researchers to help us catalyze a new generation of tools that can find the proverbial 'needle in a haystack' within data environments of unprecedented scale and complexity."

Competition Structure

The competition is structured in two rounds:

Round 1:

Detection | 2 months -- Focuses on the accurate identification and characterization of pathogens within complex environmental samples.

Round 2:

Attribution | 1.5 months -- Challenges participants to determine the origin of engineered pathogens by identifying unique physical, chemical, or design signatures.

Prizes and Opportunities

A total of \$180,000 in monetary prizes will be awarded to the top three performers in each round:

Round 1

1st place

\$50,000

2nd place

\$30,000

3rd place

\$10,000

Round 2

1st place

\$50,000

2nd place

\$30,000

3rd place

\$10,000

In addition to monetary prizes, the Bio Attribution Challenge offers significant opportunities for participants, including:

The chance to present their work to lea\,\ldots
\end{quote}

\subsection*{Federal Register (US)}
\paragraph{Low-AI example (2026, \texttt{fraction\_ai}~$=$~0.00).}

\begin{quote}\small
[Federal Register Volume 91, Number 16 (Monday, January 26, 2026)]

[Rules and Regulations]

[Pages 3048-3056]

From the Federal Register Online via the Government Publishing Office [www.gpo.gov]

[FR Doc No: 2026-01413]

-----------------------------------------------------------------------

ENVIRONMENTAL PROTECTION AGENCY

40 CFR Part 52

[EPA-R08-OAR-2024-0607; FRL-12598-02-R8]

Air Plan Disapproval; Colorado; Regional Haze Plan for the Second 

Implementation Period

AGENCY: Environmental Protection Agency (EPA).

ACTION: Final rule.

-----------------------------------------------------------------------

SUMMARY: The Environmental Protection Agency (EPA) is disapproving a 

regional haze state implementation plan (SIP) revision submitted in 

2022 by the State of Colorado under the Clean Air Act (CAA or Act) and 

the EPA's Regional Haze Rule (RHR) for the program's second 

implementation period. Colorado's 2022 regional haze SIP revision 

addresses the requirement that states revise their long-term strategies 

every implementation period to make reasonable progress towards the 

national goal of preventing any future, and remedying any existing, 

anthropogenic impairment of visibility, including regional haze, in 

mandatory Class I Federal areas (Class I areas). We are disapproving 

Colorado's 2022 regional haze SIP revision pursuant to the CAA and 

regulatory regional haze requirements. The EPA is not taking final 

action at this time on a separate revision to Colorado's SIP that 

consolidates existing, previously approved regional haze provisions 

into the same regulation where Colorado's new, second planning period 

provisions are located.

DATES: This rule is effective on February 25, 2026.

ADDRESSES: The EPA has established a docket for this action under 

Docket ID No. EPA-R08-OAR-2024-0607. All documents in the docket are 

listed on the https://www.regulations.gov website. Although listed in 

the index, some information is not publicly available, e.g., CBI o\,\ldots
\end{quote}

\paragraph{High-AI example (2026, \texttt{fraction\_ai}~$=$~0.51).}

\begin{quote}\small
[Federal Register Volume 91, Number 64 (Friday, April 3, 2026)]

[Notices]

[Pages 16962-16963]

From the Federal Register Online via the Government Publishing Office [www.gpo.gov]

[FR Doc No: 2026-06549]

-----------------------------------------------------------------------

DEPARTMENT OF HEALTH AND HUMAN SERVICES

Office of the Secretary

Statement of Organization, Functions, and Delegations of 

Authority

AGENCY: Office of the Secretary, Department of Health and Human 

Services (HHS).

ACTION: Notice.

-----------------------------------------------------------------------

SUMMARY: The Department of Health and Human Services (HHS) is issuing 

this notice to revise its Statement of Organization, Functions, and 

Delegations of Authority for the Office of the Secretary (OS). This 

reorganization removes the Office of the Chief Information Officer 

(OCIO) from the organizational description for the Office of the 

Assistant Secretary for Administration (ASA), and establishes the OCIO 

as a stand-alone organization that reports directly to the Secretary 

and Deputy Secretary. These changes supersede the OCIO-related 

organizational language contained in the notice published at 74 FR 

57747 (November 9, 2009) (document number E9-26963) and any subsequent 

amendments, as well as corresponding OCIO references in the Assistant 

Secretary for Administration Federal Register notice published at 90 FR 

3655 (January 10, 2025) (document number 2025-00382).

DATES: This reorganization is effective upon date of publication of 

this notice in the Federal Register.

FOR FURTHER INFORMATION CONTACT: Bobby D. Flanders, Jr., Office of the 

Chief Information Officer, Department of Health and Human Services, 200 

Independence Avenue SW, Washington, DC 20201, telephone: 202-969-3622, 

email: bobby.flanders@hhs.gov.

SUPPLEMENTARY INFORMATION:

I. Office of the Assistant Secretary for Administration (ASA)

    Under the heading ``Office of the Assistant Secretary for 

Administration''\,\ldots
\end{quote}

\subsection*{Military Review (US)}
\paragraph{Low-AI example (2025, \texttt{fraction\_ai}~$=$~0.00).}

\begin{quote}\small
Toggle navigation

Military Review

Current Edition

Print Archives

Digital Publications

Current Online Exclusives

Online Exclusive Archive

History of

Military Review

Creative Kiosk

Director's Select Articles

Future Warfare Writing Program

Book Reviews

Policies and Submission Guidelines

Plagiarism Policy

MR Submission Guidelines

Book Review Submission Guidelines

Future Warfare Writing Program Submission Guidelines

War Poetry Submission Guidelines

DePuy Writing Contest

Research

Special Topics

Guide to Find Past Editions

About

Contact Military Review

About Military Review

Subscribe to Military Review

Publishing Disclaimer:

In all of its publications and products,

Military Review

presents professional information. However, the views expressed therein are those of the authors and are not necessarily those of the Army University, the Department of the Army, or any other agency of the U.S. government.

The Unlikely War Hero

A Vietnam War POW's Story of Courage and Resilience in the Hanoi Hilton

Marc Leepson, Stackpole Books, 2024, 240 pages

Download the PDF

Lt. Col. Rick Baillergeon, U.S. Army, Retired

Some military history authors simply possess the aptitude to select quality book subjects. These could be topics that are interesting to the public, center on a completely overlooked event or person in history, or key on an event or person deserving of far more attention from the public. One author who clearly owns this ability is Marc Leepson. During his illustrious career, he has crafted many books that embody the above characteristics. His latest book,

The Unlikely War Hero

, focuses on a subject that is incredibly interesting, has been clearly overlooked in the past, and demands to be known by the public for his extraordinary contributions to our Nation and the Armed Forces.

Leepson's journey from selecting a subject to completed book was a long one. Over twenty-five years ago, he read an article regarding the Vietnam prisoner-of-war (POW) story of Doug Hegdahl. The author\,\ldots
\end{quote}

\paragraph{High-AI example (2024, \texttt{fraction\_ai}~$=$~0.94).}

\begin{quote}\small
Toggle navigation

Military Review

Current Edition

Print Archives

Digital Publications

Current Online Exclusives

Online Exclusive Archive

History of

Military Review

Creative Kiosk

Director's Select Articles

Future Warfare Writing Program

Book Reviews

Policies and Submission Guidelines

Plagiarism Policy

MR Submission Guidelines

Book Review Submission Guidelines

Future Warfare Writing Program Submission Guidelines

War Poetry Submission Guidelines

DePuy Writing Contest

Research

Special Topics

Guide to Find Past Editions

About

Contact Military Review

About Military Review

Subscribe to Military Review

Publishing Disclaimer:

In all of its publications and products,

Military Review

presents professional information. However, the views expressed therein are those of the authors and are not necessarily those of the Army University, the Department of the Army, or any other agency of the U.S. government.

Lead Climbers

Noncommissioned Officers Drive Change in the 10th Mountain Division

Command Sgt. Maj. Alexander D. King, U.S. Army

Download the PDF

A 10th Mountain Division soldier receives her Mountain tab during a New Soldiers Patching Ceremony 12 March 2024. (Photo courtesy of the U.S. Army)

In the heart of Fort Drum, New York, a ceremony unfolds that is as timeless as it is inspiring. Rows of new soldiers, their faces alight with pride and anticipation, stand at attention. Each one proudly bears the powder keg patch of the 10th Mountain Division, a symbol of their entry into a lineage steeped in valor and resilience, a lineage that has been shaped and guarded by the division's seasoned noncommissioned officers (NCO). From platoon sergeants to the command sergeants major and first sergeants, these NCOs have been the architects of the division's past, present, and future. They have sung the 10th Mountain Division song, a powerful anthem that bridges generations and cements a shared commitment to the Nation and to the division's alpine heritage. This moment encapsulates a profou\,\ldots
\end{quote}

\paragraph{High-AI example (2025, \texttt{fraction\_ai}~$=$~0.53).}

\begin{quote}\small
Toggle navigation

Military Review

Current Edition

Print Archives

Digital Publications

Current Online Exclusives

Online Exclusive Archive

History of

Military Review

Creative Kiosk

Director's Select Articles

Future Warfare Writing Program

Book Reviews

Policies and Submission Guidelines

Plagiarism Policy

MR Submission Guidelines

Book Review Submission Guidelines

Future Warfare Writing Program Submission Guidelines

War Poetry Submission Guidelines

DePuy Writing Contest

Research

Special Topics

Guide to Find Past Editions

About

Contact Military Review

About Military Review

Subscribe to Military Review

Publishing Disclaimer:

In all of its publications and products,

Military Review

presents professional information. However, the views expressed therein are those of the authors and are not necessarily those of the Army University, the Department of the Army, or any other agency of the U.S. government.

Leveraging Flexible Partnerships and the Thucydides Dance

Rethinking U.S. Foreign Policy in the Indo-Pacific Command

Lt. Col. Patrick O. Boling, PhD, Louisiana National Guard

Paul R. Sanders, PhD

Download the PDF

Soldiers assigned to Company B, 1st Battalion, 5th Infantry Regiment, 1st Infantry Brigade Combat Team, 11th Airborne Division, train with Indian Army soldiers assigned to the 4/8 Gorkha Rifles Infantry Battalion, 91st Infantry Brigade, during Exercise Tiger Triumph near Visakhapatnam, India, on 4 April 2025. Tiger Triumph is a joint and combined U.S.-India exercise focusing on humanitarian assistance, disaster response readiness, and interoperability in the Indian Ocean region and beyond to support a free and open Indo-Pacific. (Photo courtesy of the Indian Army)

At dawn Sun Pin lured P'ang Ch?an and half his army onto a narrow path along which Sun Pin had removed bark from a large tree. Sun Pin positioned his army in ambush along the trail with the instruction to fire when they saw a torch. General P'ang Ch?an was summoned to the bare tree by his advanced guard. He l\,\ldots
\end{quote}

\subsection*{OSTP (US)}
\paragraph{Low-AI example (2021, \texttt{fraction\_ai}~$=$~0.00).}

\begin{quote}\small
Remarks by Dr. Eric Lander to The World Academy of Sciences

Members of the World Academy of Sciences, it is an honor to join you. I especially want to thank Academy President Dr. Mohamed Hassan, and Executive Director Dr. Romain Murenzi --- both for their invitation, and for their recent work to help rescue and resettle refugee scholars from Afghanistan.

Today I'd like to talk about the values we share as scientists: the values of curiosity, openness, humility, diversity, and dissent.

We value curiosity, because it's the bedrock of scientific discovery.

We value openness, because science doesn't move forward unless we share our knowledge and breakthroughs so that others can build on them.

We value humility, because no one has a monopoly on the best ideas. And despite what others may think, great science doesn't only happen at elite institutions, in wealthy contexts, or western countries. It happens all around the world.

We value diversity, because scientific progress depends on someone seeing questions or answers that no one has seen before --- because they bring a different lens, different experiences, different questions, different passions.

And we value dissent, because progress in science depends on challenging ideas and challenging theories to see if they stand up to scrutiny.

These values are what make science such an effective force for progress. They mean that a graduate student can have a better idea than the most distinguished professor, and should be recognized for that. They make science, at its best, one of the world's most democratic practices.

In most global contexts, countries do not meet as equals. But here in science, we should all strive to meet as equals. Because scientific insights and breakthroughs can come from anywhere, and anybody.

Now, look, I'm a realist: I know that science sometimes fails to live up to these values --- in individual labs, at scientific journals, in scientific institutions. I know that some countries, including my own, sometime\,\ldots
\end{quote}

\end{document}